\documentclass[fleqn,11pt]{wlscirep}
\usepackage[utf8]{inputenc}
\usepackage[T1]{fontenc}
\usepackage{microtype}
\DisableLigatures{encoding = T1, family = tt* }
\usepackage[english]{babel}
\usepackage{graphicx}
\usepackage{amsmath}
\usepackage{amssymb}
\usepackage{amsfonts}
\usepackage{wasysym}
\usepackage[vcentermath]{youngtab}
\usepackage{upgreek}
\usepackage{bm,dsfont}
\usepackage{multirow}
\usepackage{enumitem}
\usepackage{makecell}
\usepackage{framed}
\usepackage{color}
\usepackage{hyperref}
\hypersetup{
colorlinks = true,
linkcolor = [rgb]{0.70,0.13,0.13},
citecolor = [rgb]{0.13,0.55,0.13},
urlcolor = [rgb]{0.25, 0.41, 0.88}}

\newcommand{\bra}[1]{\langle #1|}
\newcommand{\ket}[1]{|#1 \rangle}

\newcommand{\ii}{\mathrm{i}}

\newcommand{\id}{\mathds{1}}
\newcommand{\E}{\mathop{\mathbb{E}}}

\renewcommand{\O}{\mathrm{O}}

\renewcommand{\o}{\mathfrak{o}}

\newcommand{\dsE}{\mathbb{E}}

\newcommand{\scL}{\mathcal{L}}

\newcommand{\scN}{\mathcal{N}}

\newcommand{\Tr}{\operatorname{Tr}}

\newcommand{\vect}[1]{{\bm{#1}}}
\newcommand{\ttt}[1]{{\text{\texttt{#1}}}}

\newcommand{\eq}[1]{\begin{equation}#1\end{equation}}
\newcommand{\eqs}[1]{\begin{equation}\begin{split}#1\end{split}\end{equation}}
\newcommand{\eqnref}[1]{Eq.\,\eqref{#1}}
\newcommand{\figref}[1]{Fig.\,\ref{#1}}
\newcommand{\tabref}[1]{Tab.\,\ref{#1}}

\title{Observing Schrödinger's Cat with Artificial Intelligence: Emergent Classicality from Information Bottleneck}

\author[1,2]{Zhelun Zhang}
\author[3,*]{Yi-Zhuang You}
\affil[1]{School of Physics, Peking University, Beijing 100871, China}
\affil[2]{Department of Physics, Harvard University, Cambridge, MA 02138, USA}
\affil[3]{Department of Physics, University of California, San Diego, CA 92093, USA}

\affil[*]{Corresponding author: yzyou@physics.ucsd.edu}

\keywords{Quantum Mechanics, Emergent Classicality, Information Bottleneck}

\begin{abstract}
We train a generative language model on the randomized local measurement data collected from Schrödinger’s cat quantum state. We demonstrate that the classical reality emerges in the language model due to the information bottleneck: although our training data contains the full quantum information about Schrödinger’s cat, a weak language model can only learn to capture the classical reality of the cat from the data. We identify the quantum-classical boundary in terms of both the size of the quantum system and the information processing power of the classical intelligent agent, which indicates that a stronger agent can realize more quantum nature in the environmental noise surrounding the quantum system. Our approach opens up a new avenue for using the big data generated on noisy intermediate-scale quantum (NISQ) devices to train generative models for representation learning of quantum operators, which might be a step toward our ultimate goal of creating an artificial intelligence quantum physicist.
\end{abstract}

\begin{document}

\flushbottom
\maketitle
%
%
\thispagestyle{empty}


\section*{Introduction}

Quantum mechanics offers a remarkably precise depiction of nature at its most fundamental level, particularly in the world of microscopic particles where phenomena like quantum uncertainty, coherence, and entanglement prevail. Yet, our everyday experiences are firmly anchored in the classical world, where macroscopic objects follow well-defined trajectories in a deterministic manner, and the peculiarities of quantum behavior seem imperceptible. This discrepancy between the quantum and classical realms presents a profound enigma in theoretical physics: the quantum-to-classical transition \cite{Leggett2002T, Schlosshauer2014T1404.2635}, or how and why the classical world emerges from the underlying quantum reality. 

Historically, this enigma was epitomized by the paradox of Schrödinger's cat\cite{Schrodinger1935D} --- a thought experiment in which a hypothetical cat can be prepared in a quantum superposition state of both alive and dead, although we have never witnessed such a superposition cat in our daily life. According to the Copenhagen interpretation, the act of observing the cat triggers a collapse of its superposition state into one of the two classical realities: either the cat is alive or it is dead. However, this explanation raises further questions about the role of the observer and the nature of quantum state collapse. Over the years, many theories have been proposed to better understand the emergence of classicality in quantum many-body systems, including decoherence theory \cite{Zeh1970O, Joos1985Emergence, Schlosshauer2004Dquant-ph/0312059}, quantum Darwinism \cite{Zurek1981Pointer, Zurek1982Environment-induced, Zurek1993Preferred, Zurek1998Decoherence, Zurek2009Quantum}, many-worlds interpretation \cite{Everett1957R, Everett2015T, Tegmark1998Tquant-ph/9709032}, spontaneous localization \cite{Ghirardi1986U, Ghirardi1990M, Bassi2013Models}, quantum Bayesianism \cite{Home1992E, Fuchs2001Qquant-ph/0106166, Caves2002Qquant-ph/0106133, Fuchs2002Qquant-ph/0205039, Caves2002Cquant-ph/0206110, Fuchs2004U, Fuchs2013Q1301.3274, Mermin2014P}, and information-based interpretations \cite{Brandao2015G1310.8640, Foti2019W1810.10261, Qi2021E2001.01507, Coppo2022T2203.13587}. A consistent modern understanding is gradually crystallizing from these diverse perspectives.

Decoherence provides a key mechanism bridging the quantum and classical worlds. It arises from the inevitable interaction of a quantum system with its environment, causing the ``leaking'' of quantum information into the surroundings and the subsequent loss of quantum coherence. Spontaneous localization suggests that the effects of decoherence can be modeled as spontaneous \emph{random local measurements} of the quantum system by the environment. These measurements extract classical information about the quantum system and spread them in the environment. Quantum Darwinism further explains the quantum state collapse as a result of the natural selection of a classical reality that is consistent with the classical information proliferated in the environment. This perspective aligns with quantum Bayesianism, which interprets quantum states as descriptions of beliefs and expectations regarding potential future experimental outcomes. The classical reality, in this view, emerges as an intelligent agent updates its belief based on the observed randomized measurement outcomes in the environment.

It is conceivable that an agent's ability to process classical information could influence its interpretation of reality. This task of reconstructing quantum states from classical information is referred to as \emph{quantum state tomography} \cite{Paris2004Q, Cramer2010E1101.4366, Flammia2012Q1205.2300, ODonnell2015E1508.01907, Haah2015S1508.01797, 
Lanyon2017E1612.08000, Brandao2017Q1710.02581, Aaronson2017S1711.01053, Wang2017S1712.03213, Aaronson2019G1904.08747} in quantum information science. If the environment can provide classical descriptions of sufficiently many copies of identical quantum states in different measurement basis, an agent with powerful enough classical information processing abilities could, in theory, reconstruct the full quantum reality from the classical data with considerable accuracy. This principle has been demonstrated in research on quantum state tomography, especially in recent advances of \emph{classical shadow tomography} \cite{Ohliger2013E1204.5735, Guta2018F1809.11162, Huang2020P2002.08953}. We hypothesize that the difficulty we often experience in comprehending the full quantum reality as compared to the classical reality might be linked to our limited ability in processing classical information.

\begin{figure}[htbp]
\begin{center}
\includegraphics[scale=0.7]{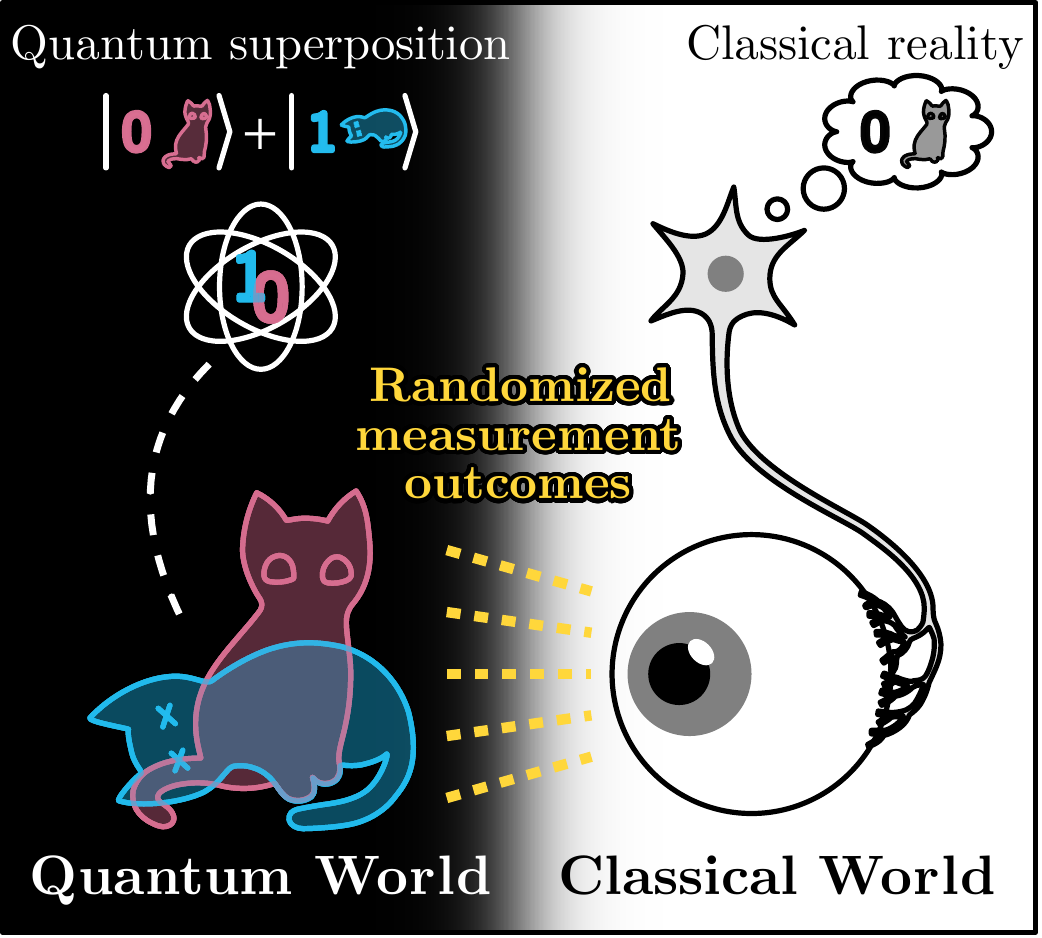}
\caption{Illustration of the general idea. Quantum evolution prepares an entangled Schrödinger's cat state in the quantum world. Decoherence occurs as random local measurements by the environment, which serves as the quantum-classical interface. The randomized measurement outcomes train an intelligent agent in the classical world, such that the agent can realize and identify the emergent classical reality.}
\label{fig: cartoon}
\end{center}
\end{figure}

To test this hypothesis, we propose training a generative language model\cite{Radford2018I} on random local measurement outcomes gathered from Schrödinger's cat quantum state. The trained model can then be prompted with new experiment proposals to explore its understanding of the reality of Schrödinger's cat, thereby investigating the emergent classicality from the perspective of artificial intelligence. \figref{fig: cartoon} provides a cartoon illustration of our setup. In this research, we do not intend to address how the quantum state collapses under the randomized measurements from the environment. Instead, we will adhere to the standard quantum mechanical approach to simulate the randomized measurement outcomes that the environment could collect. Our primary question is to what extent a classical intelligent agent (or a classical algorithm) can process this classical information to form an understanding of reality. More importantly, we seek to study how this emergent reality is influenced by the size of the quantum system and the information bottleneck\cite{Tishby2000Tphysics/0004057, Tishby2015D1503.02406} of the classical agent. Through this research, we hope to quantitatively identify the boundary between the quantum and classical worlds\cite{Fisher2015Q1508.05929}, should one exist.

\section*{Methods}

\subsection*{Randomized Measurement Scheme}

\begin{figure}[htbp]
\begin{center}
\includegraphics[scale=0.7]{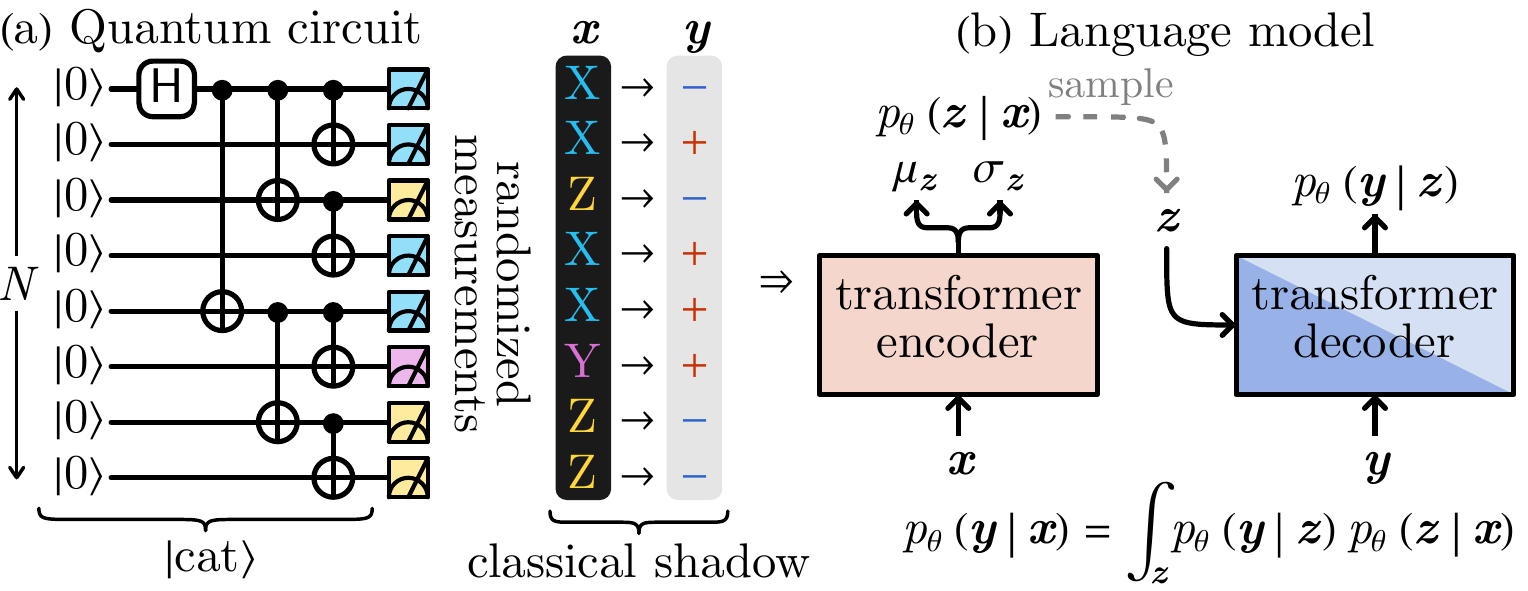}
\caption{The model setup. (a) The quantum circuit prepares the cat state. The random local measurements collapse the cat state and generates the classical shadow $(\vect{x},\vect{y})$. (b) The classical shadow data is used to train a generative language model for $p(\vect{y}|\vect{x})$, built with a transformer-based $\beta$-VAE architecture.}
\label{fig: setup}
\end{center}
\end{figure}

We begin with an $N$-qubit Greenberger-Horne-Zeilinger (GHZ) state\cite{Greenberger2007G0712.0921} as a model for the quantum state of Schrödinger's cat, denoted as
\eq{\ket{\text{cat}}=\frac{\ket{0}^{\otimes N}+\ket{1}^{\otimes N}}{\sqrt{2}}.}
This state can be repeatedly prepared\cite{Raimond2001M} by a quantum circuit depicted in \figref{fig: setup}(a), which comprises a Hadamard gate followed by a series of controlled-NOT gates\cite{Nielsen2000Q}. This circuit mimics the unitary quantum dynamics that generates Schrödinger's cat state by entangling the qubits together.

The decoherence of Schrödinger's cat in the environment can be simulated by a series of random local measurements, which represent the environment's random interactions with the cat, akin to events such as air molecules bouncing off the body of the cat. While we could assume these measurements to be weak and continuous for a more accurate reflection of reality, this assumption is not essential for our discussion. For simplicity, we assume that the environment randomly selects one of the three Pauli observables $\{X,Y,Z\}$ for each qubit and performs a projective measurement of the chosen Pauli observable. As a result, the cat state will collapse into certain post-measurement state. We will not delve into the nature of how this process occurs, as it is not the focus of our study. We merely follow the principles of quantum mechanics to simulate the measurement process and collect the binary measurement outcomes $\{\pm 1\}$. We regard these outcomes as the classical information dispersed in the environment after the decoherence of the cat. Our goal is to analyze how much we can tell about the original quantum state from such classical information.

\subsection*{Classical Shadow Data Structure}

Specifically, the data from each random measurement can be represented as a pair of sequences denoted as $(\vect{x},\vect{y})$, where $\vect{x}\in\{X,Y,Z\}^{\times N}$ is the observable sequence and $\vect{y}\in\{\pm1\}^{\times N}$ is the measurement outcome sequence, as exemplified in \figref{fig: setup}. Both are sequences of $N$ tokens. Their joint probability distribution, $p_\text{dat}(\vect{x},\vect{y})=p(\vect{y}|\vect{x})p(\vect{x})$, defines the \emph{data distribution}, where $p(\vect{x})=3^{-N}$ is the probability of randomly choosing an observable sequence $\vect{x}$, which is assumed to be uniform, and
\eq{\label{eq: p(y|x) dat} 
p(\vect{y}|\vect{x})=\bra{\text{cat}}\bigotimes_i\frac{1+y_ix_i}{2}\ket{\text{cat}}} is the probability for the measurement outcomes $\vect{y}$ to occur, which is calculated according to Born's rule in quantum mechanics. It encodes non-trivial information about the original quantum state $\ket{\text{cat}}$.

We build a classical simulator to sample the sequence pair $(\vect{x},\vect{y})$ from the distribution $p_\text{dat}(\vect{x},\vect{y})$ upon request. This essentially simulates the repeated process of creating the Schrödinger's cat state, allowing it to decohere, and collecting the classical information it leaves behind in the environment. Example samples of $(\vect{x},\vect{y})$ sequence pairs can be found in Supplementary Information. These $(\vect{x},\vect{y})$ sequence pairs, also referred to as \emph{classical shadows} of the original quantum state, describe random projections of the quantum state in a random measurement basis, akin to a high-dimensional object casting a shadow in a low-dimensional subspace. Classical shadow tomography offers a systematic classical post-processing technique for quantum state reconstruction from its classical shadows\cite{Ohliger2013E1204.5735, Guta2018F1809.11162, Huang2020P2002.08953}. Given the randomized Pauli measurement scheme mentioned above, the reconstruction formula is
\eq{\label{eq: rec cat}
\rho_\text{cat}:=\ket{\text{cat}}\bra{\text{cat}}=\mathop{\dsE}_{(\vect{x},\vect{y})\sim p_\text{dat}}\bigotimes_i\frac{1+3y_ix_i}{2}.}
This demonstrates that given a sufficient amount of classical data about repeated copies of a quantum state, it is in principle possible to accurately reconstruct the full quantum reality.

\subsection*{Generative Modeling of Classical Shadows}

If we are short of memory resources to store the entire dataset of classical shadows, a potential workaround is to train a generative model ``on the fly'' as we collect the classical shadow data. Once trained, the generative model can approximate the data distribution $p_\text{dat}(\vect{x},\vect{y})$ with a model distribution $p_\text{mdl}(\vect{x},\vect{y})$ and provide us with an endless supply of samples. This approach enables a more efficient compression and utilization of the classical shadow data, gaining an edge in addressing quantum problems. Many recent studies \cite{Li2020V2012.08288, Huang2021P2011.01938, Huang2021I2101.02464, Huang2021P2106.12627, Huang2022Q2112.00778, Van-Kirk2022H2212.06084, Wei2023N2305.01078, Jerbi2023S2306.00061} have demonstrated the theoretical and practical advantages of combining machine learning with classical shadows.

In constructing the probability model $p_\text{mdl}(\vect{x},\vect{y})=p_\theta(\vect{y}|\vect{x})p(\vect{x})$, our focus lies in modeling the conditional distribution $p(\vect{y}|\vect{x})$ with parameters $\theta$. This is because $p(\vect{x})=3^{-N}$ is a trivial uniform distribution that does not need modeling. If we perceive the observable sequence $\vect{x}$ as a question, and the measurement outcome sequence $\vect{y}$ as an answer to that question by the quantum experiment, then the modeling of $p(\vect{y}|\vect{x})$ can be formulated as a \emph{chat completion} task in natural language processing, which suggests the generative language model as a natural solution. Once trained, the language model can take over the role of the quantum experiment to answer inquiries about the underlying quantum state $\ket{\text{cat}}$. In other words, the model can ``speak'' the quantum language. The learning process imitates the way an intelligent agent accumulates knowledge about the world by observing the environment.

The transformer\cite{Vaswani2017A1706.03762} architecture stands out as a natural choice for modeling $p(\vect{y}|\vect{x})$. As illustrated in \figref{fig: setup}(b), we have made a slight modification in its latent space by imposing a variational information bottleneck \cite{Tishby2000Tphysics/0004057, Tishby2015D1503.02406} borrowed from the $\beta$-variational auto-encoder ($\beta$-VAE)\cite{Higgins2017b} architecture. This structure allows us to adjust the model's information processing power, which will be crucial for our subsequent study. The transformer-based $\beta$-VAE  comprises two probability models: an encoder $p_\theta(\vect{z}|\vect{x})$ that infers latent variables $\vect{z}$ from the input sequence $\vect{x}$, and a decoder $p_\theta(\vect{y}|\vect{z})$ that generates the output sequence $\vect{y}$ based on $\vect{z}$, such that
\eq{\label{eq: p(y|x) mdl}
p_\theta(\vect{y}|\vect{x})=\int_\vect{z} p_\theta(\vect{y}|\vect{z})p_\theta(\vect{z}|\vect{x}).} A more detailed description of the architecture can be found in the Supplementary Information. The goal is to approximate $p(\vect{y}|\vect{x})$ in \eqnref{eq: p(y|x) dat} with $p_\theta(\vect{y}|\vect{x})$ in \eqnref{eq: p(y|x) mdl} by optimizing the model parameters $\theta$.

The model can be trained by minimizing the $\beta$-VAE loss $\scL=\mathop{\dsE}_{(\vect{x},\vect{y})\sim p_\text{dat}}\scL(\vect{x},\vect{y})$ on the training data of classical shadows collected from the cat state, where the loss function for each classical shadow $(\vect{x},\vect{y})$ reads 
\eq{\scL(\vect{x},\vect{y})=-\E_{\vect{z}\sim p_\theta(\vect{z}|\vect{x})}\log p_\theta(\vect{y}|\vect{z})+\beta D_\text{KL}[p_\theta(\vect{z}|\vect{x})\Vert p_\scN(\vect{z})].}
The first term is the negative log likelihood loss and the second term is a Kullback-Leibler (KL) divergence regularization. $p_\scN(\vect{z})$ denotes the normal distribution of zero mean and unit variance. The hyper-parameter $\beta$ permits us to adjust the variational information bottleneck of the transformer. A large $\beta$ enforces $p_\theta(\vect{z}|\vect{x})$ to approach $p_\scN(\vect{z})$ regardless of $\vect{x}$, which limits the model's ability to encode information about $\vect{x}$  in the latent variables $\vect{z}$. Therefore, increasing the hyperparameter $\beta$ will impose a stronger information bottleneck, thereby diminishing the model’s information processing capacity.

\section*{Results}

\subsection*{Model Evaluation}

We take an $N$-qubit cat state, collect its classical shadows, and train a generative language model concurrently. For each level of the information bottleneck strength $\beta$ and each distinct qubit number $N$, we train a separate model. Upon convergence of the training, we evaluate the performance of each model as follows. First, we sample from the model distribution $p_\text{mdl}(\vect{x},\vect{y})$ by prompting the model with a random observable sequence $\vect{x}$ and collect the model generated measurement outcome sequence $\vect{y}$. Then, we use the classical shadow tomography approach to reconstruct a quantum state $\rho_\text{mdl}$ based on the model generated classical shadows,
\eq{\label{eq: rec mdl}
\rho_\text{mdl}=\mathop{\dsE}_{(\vect{x},\vect{y})\sim p_\text{mdl}}\bigotimes_i\frac{1+3y_ix_i}{2}.}
Finally, we evaluate the model constructed quantum state $\rho_\text{mdl}$ by two metrics:
\begin{itemize}
\item Quantum fidelity: $F(\rho_\text{cat}, \rho_\text{mdl})=\bra{\text{cat}}\rho_\text{mdl}\ket{\text{cat}}$, given that $\rho_\text{cat}=\ket{\text{cat}}\bra{\text{cat}}$ is a pure state. The fidelity measures how closely the state $\rho_\text{mdl}$ approximates the original cat state.
\item Von Neumann entropy: $S(\rho_\text{mdl})=-\Tr \rho_\text{mdl}\log\rho_\text{mdl}$. The entropy quantifies the disorder or uncertainty of a quantum state.  A zero entropy indicates that $\rho_\text{mdl}$ is pure.
\end{itemize}
If the model is strong enough to reconstruct the full quantum reality, i.e., $\rho_\text{mdl}=\rho_\text{cat}$, we should expect the fidelity $F(\rho_\text{cat}, \rho_\text{mdl})=1$ to be one and the entropy $S(\rho_\text{mdl})=0$ to be zero.

\begin{figure}[htbp]
\begin{center}
\includegraphics[scale=0.7]{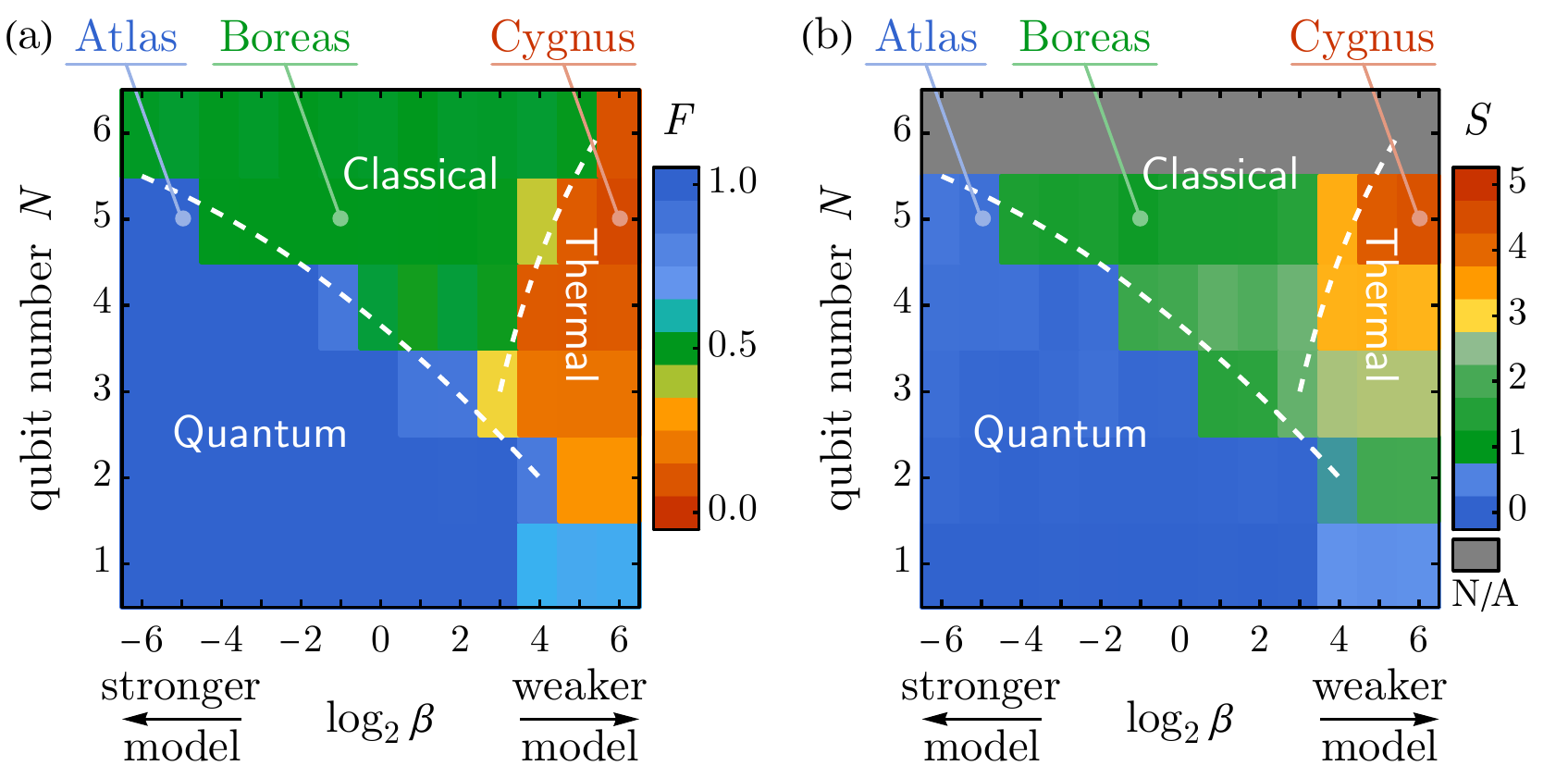}
\caption{(a) Quantum fidelity and (b) von Neumann entropy of the model reconstructed state $\rho_\text{mdl}$ for model trained at different $\beta$ (in logarithmic scale) and $N$. Dashed curves are suggestive cross-over boundaries. The entropy evaluation for $N=6$ is not available, as we are not aware of an efficient approach to estimate entropy other than the full state tomography (which becomes computationally infeasible for $N=6$). Three representative models are named as Atlas, Boreas and Cygnus.}
\label{fig: phase}
\end{center}
\end{figure}

\figref{fig: phase} presents fidelity and entropy evaluations for various models. When $\beta$ is small, the model reconstructed state $\rho_\text{mdl}$ approximates the cat state $\rho_\text{cat}$, as indicated from $F(\rho_\text{cat}, \rho_\text{mdl})\approx 1$ and $S(\rho_\text{mdl}) \approx 0$. This suggests that the model has learnt the complete quantum reality from the classical shadows. We label this parameter region as the ``quantum'' regime. Away from this regime, the quality of $\rho_\text{mdl}$ deteriorates as $\beta$ increases. This is due to the model's declining ability to capture the statistical features of the classical shadows under a more restrictive information bottleneck. Eventually, for large $\beta$, the model generates $(\vect{x},\vect{y})$ almost uniformly, corresponding to a maximally mixed state $\rho_\text{mdl}\simeq \id /2^N$ roughly. We mark this limit as ``thermal''. Interestingly, as the qubit number $N$ increases, an intermediate ``classical'' regime emerges. In this regime, the reconstructed state $\rho_\text{mdl}\simeq \frac{1}{2}(\ket{0}^{\otimes N}\bra{0}^{\otimes N}+\ket{1}^{\otimes N}\bra{1}^{\otimes N})$ is approximately the decohered density matrix, signifying that the model has learnt the distinct classical realities of Schrödinger's cat but is unable to discern the quantum coherence. 

To justify the above interpretations, we selected three representative models from these three regimes separately, named Atlas, Boreas, and Cygnus (standing for A, B, C, respectively). They are trained on the $N=5$ classical shadow data with different hyperparameters $\beta=2^{-5},2^{-1},2^{6}$ respectively, as marked out in \figref{fig: phase}. To understand the differences between Atlas, Boreas, and Cygnus, let us chat with them!

\subsection*{One-Shot Classification Tasks}

\begin{figure}[htbp]
\begin{center}
\includegraphics[scale=0.7]{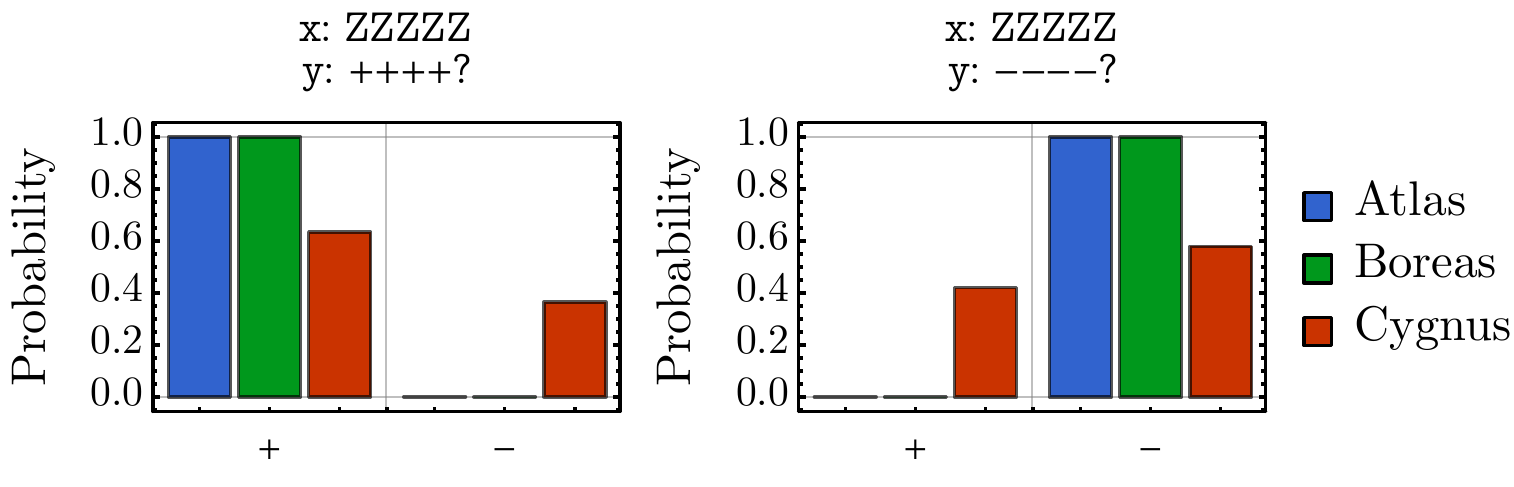}
\caption{Performances of three representative models on the one-shot cat classification task.}
\label{fig: alivedead}
\end{center}
\end{figure}

We can guide the language models to perform different classification tasks by prompt engineering. The first problem we are interested in is: given a one-shot observation of a Schrödinger’s cat, try to determine whether it is alive or dead. Here's how we might prompt the model:
\eqs{\label{eq: cat +}
\vect{x}&: \ttt{ZZZZZ}\\
\vect{y}&: \ttt{++++?}}
Here ``\texttt{?}'' stands for a blank token for the language model to complete. This is akin to asking, ``If most of the cat's cells are alive, is the cat alive or dead?'' If the model has learnt the perfect correlation among the $Z$ measurement outcomes on the cat state, it will choose to fill in the blank with a ``\texttt{+}''. Similarly, for the prompt:
\eqs{\label{eq: cat -}
\vect{x}&: \ttt{ZZZZZ}\\
\vect{y}&: \ttt{----?}}
We would expect the model to complete the sequence with a ``\texttt{-}''. Answering these questions essentially classifies the observed cat into alive and dead categories. \figref{fig: alivedead} shows the performance of Atlas, Boreas, and Cygnus in this test. We observe that Atlas and Boreas perform flawlessly on the task, while Cygnus is essentially guessing.

\begin{figure}[htbp]
\begin{center}
\includegraphics[scale=0.7]{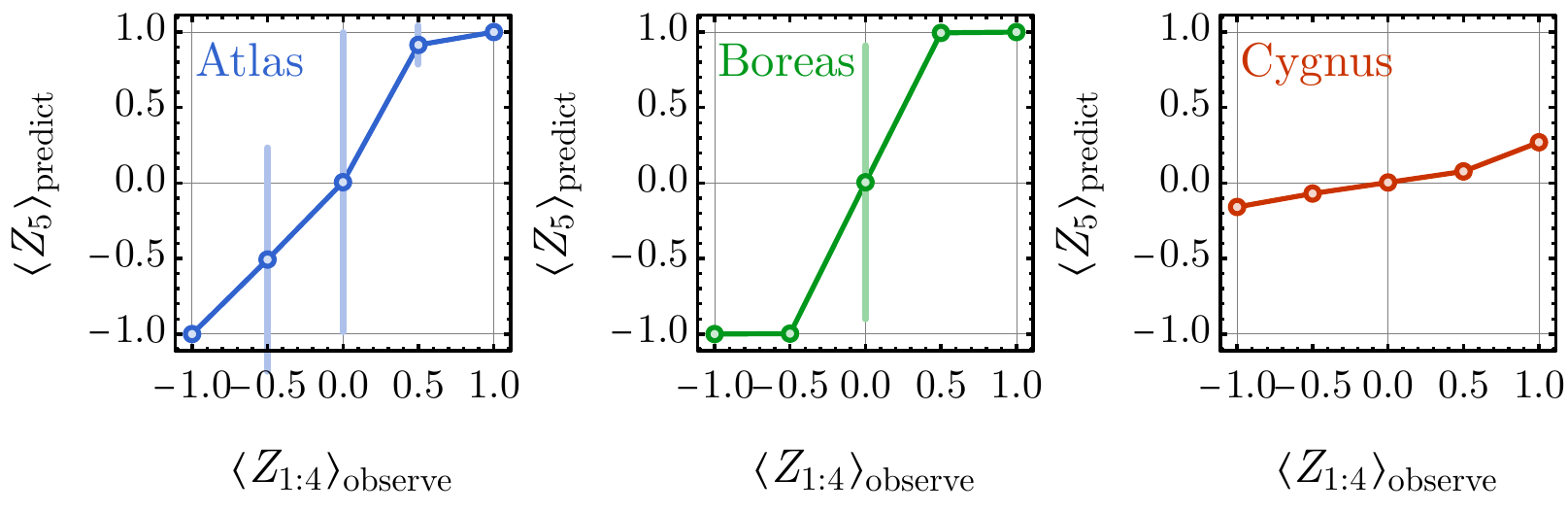}
\caption{Behaviors of three representative models under out-of-distribution prompts for the one-shot cat classification task. Error bars indicate the mean deviations.}
\label{fig: alivedead ood}
\end{center}
\end{figure}

Then what about the following prompt?
\eqs{
\vect{x}&: \ttt{ZZZZZ}\\
\vect{y}&: \ttt{+-+-?}}
This is an out-of-distribution prompt, since it will never appear as a classical shadow of the cat state due to the mismatched $Z$-basis measurement outcomes. We test the representative models will all combinations of the $Z_i$ (for $i=1,2,3,4$) measurement outcomes of the first four qubits, and collect the models' responses of $Z_5$ measurement outcomes. The predicted $Z$-polarization $\langle Z_5\rangle_\text{predict}$ is plotted against the observed average $Z$-polarization $\langle Z_{1:4}\rangle_\text{observe}:=\frac{1}{4}\sum_{i=1}^{4}\langle Z_i\rangle$ in \figref{fig: alivedead ood}. The tests at the two limits of $\langle Z_{1:4}\rangle_\text{observe}=\pm1$ belong to in-distribution tests, while the remaining tests are out-of-distribution. From our results, it appears that both Atlas and Boreas are capable of generating reasonable interpolations between the two in-distribution limits. However, Boreas seems to form a binary understanding of the cat's state, either live or dead, while Atlas exhibits a more non-binary understanding, viewing the transition from alive to dead as a continuous spectrum.

\begin{figure}[htbp]
\begin{center}
\includegraphics[scale=0.7]{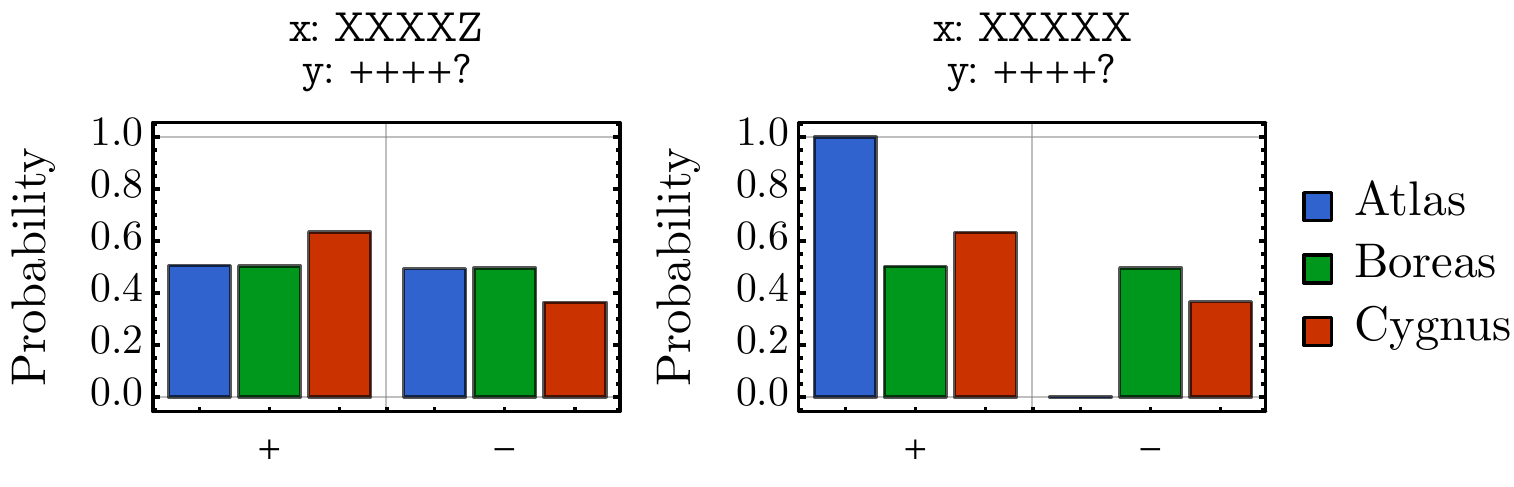}
\caption{Performances of three representative models on the one-shot cat classification and coherence prediction tasks, when the previous measurements has not collapses the superposition.}
\label{fig: coherence}
\end{center}
\end{figure}

We are also interested in whether these models can decode the quantum coherence encoded in the classical shadow data. In previous examples, local $Z$-measurements destroy the quantum coherence of the cat state, preventing us from testing coherence on the last qubit. To preserve the quantum coherence, we turn to local X-measurements. Suppose the first four measurement outcomes are $X_i=+1$ (for $i=1,2,3,4$). This prepares the last qubit into a superposition state $\frac{1}{\sqrt{2}}(\ket{0}+\ket{1})$. We can examine the models' understanding of this state using the following prompts:
\eqs{\label{eq: Z test}
\vect{x}&: \ttt{XXXXZ}\\
\vect{y}&: \ttt{++++?}}
\eqs{\label{eq: X test}
\vect{x}&: \ttt{XXXXX}\\
\vect{y}&: \ttt{++++?}}
The $Z$-test in \eqnref{eq: Z test} is like asking ``Q: Is  Schrödinger's cat alive or dead? (\texttt{+}) Alive. (\texttt{-}) Dead.'', while the $X$-test in \eqnref{eq: X test} corresponds to probing ``Q: What is the sign of quantum coherence between alive and dead? (\texttt{+}) Positive. (\texttt{-}) Negative.'' Performances of Atlas, Boreas and Cygnus are shown in \figref{fig: coherence}. While both Atlas and Boreas can realize the superimposed classical realities, only Atlas can correctly predict the quantum coherence between them.

\tabref{tab: compare} summarizes the performances of the representative models on the cat classification task \eqnref{eq: cat +} and the coherence prediction task \eqnref{eq: X test}, together with their fidelity and entropy evaluations. These results indicate that Atlas nicely captures the quantum nature of Schrödinger’s cat, Boreas exhibits a strong understanding of classical reality, while Cygnus lacks a clear grasp of reality. They  represent models in the quantum, classical and thermal regimes respectively in \figref{fig: phase}. Their reconstructed density matrices $\rho_\text{mdl}$ can be found in Supplementary Information.

\begin{table}[htp]
\caption{Quantitative comparison of three representative models.}
\begin{center}
\begin{tabular}{c|ccc}
 & Atlas & Boreas & Cygnus \\
\hline
Task \eqnref{eq: cat +} accuracy & $\bf{1.000}$ & $\bf{1.000}$ & $0.607$\\
Task \eqnref{eq: X test} accuracy & $\bf{1.000}$ & $0.503$ & $0.634$\\
\hline
$F(\rho_\text{cat},\rho_\text{mdl})$ & $\bf{1.000}$ & $0.500$ & $0.063$ \\
$S(\rho_\text{mdl})$ [bit] & $\bf{0.206}$ & $1.190$ & $4.410$  
\end{tabular}
\end{center}
\label{tab: compare}
\end{table}

\subsection*{Latent Representations of Observable Sequences}

\begin{figure}[htbp]
\begin{center}
\includegraphics[scale=0.7]{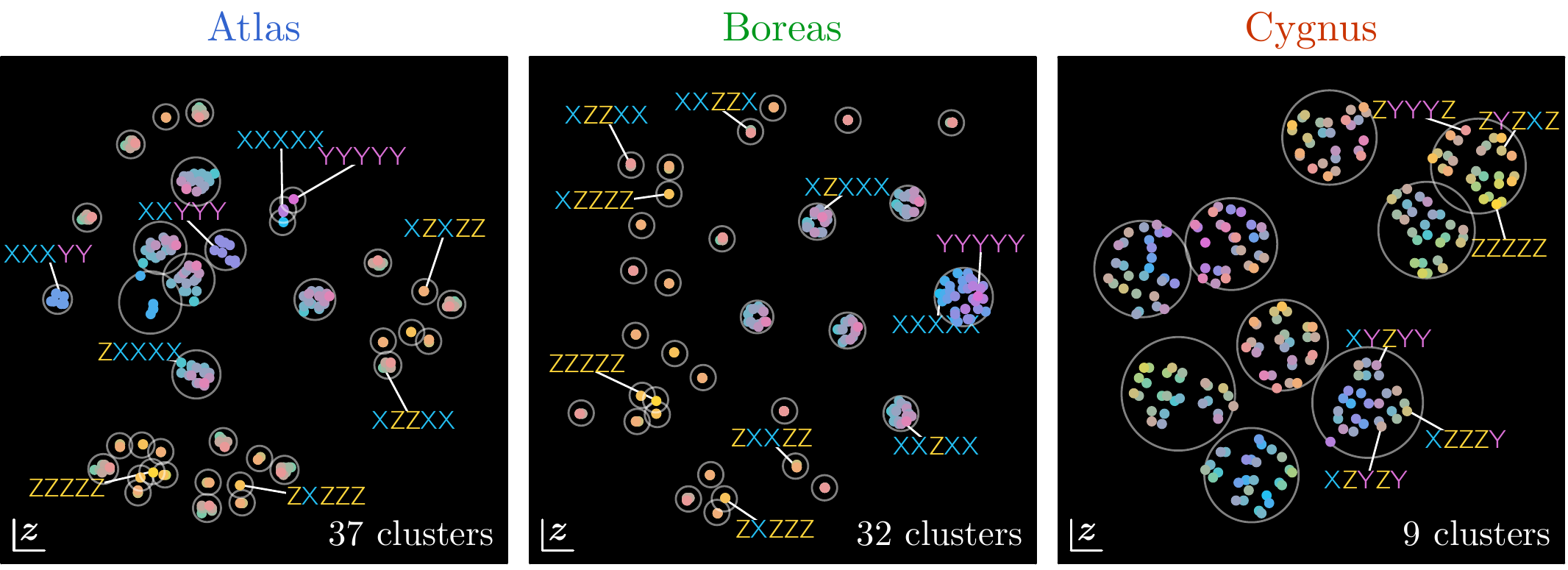}
\caption{Visualizations of latent encoding of all $3^5=243$ distinct observable sequences for three representative models. Each dot represents an observable sequence, and is colored according to the proportionality of $X$ (cyan), $Y$ (magenta), $Z$ (yellow) in the sequence. Different clusters are encircled for ease of view. }
\label{fig: starchart}
\end{center}
\end{figure}

To better understand how the information bottleneck constrains the model's ability to generate the outcome sequence $\vect{y}$ based on observable sequence $\vect{x}$, we examine how Atlas, Boreas, and Cygnus utilize the latent space to encode $\vect{x}$. \figref{fig: starchart} presents the t-SNE visualizations of the latent representations $\mu_\vect{z}(\vect{x})$ for all $\vect{x}\in\{X,Y,Z\}^{\times N}$ as inferred by different models, where $\mu_\vect{z}$ stands for the mean of the latent variables $\vect{z}$ as computed by the transformer encoder (see \figref{fig: setup}(b)). t-SNE (t-Distributed Stochastic Neighbor Embedding)\cite{Hinton2002S,van2008visualizing} is a non-linear dimensionality reduction technique, useful for visualizing high-dimensional data.

We find that the observable sequence embeddings are clustered in the latent space, and Atlas provides the most finely devided clustering. For predicting measurement outcomes, there are two important aspects about observables that the encoder should convey:
\begin{enumerate}
\item The locations of $Z$ observables. Since the measurement outcomes of $Z$ observables are all identical, the decoder needs to know where all $Z$ observables are in order to correctly correlate the measurement outcomes on these qubits.

\item The number of $Y$ observables in pure-$X/Y$ sequences. The quantum coherence of the cat state is reflected in the high-order ($N$-qubit) correlations among $X$ and $Y$ observables. Consider a string operator $S=\prod_{i=1}^{N}S_i$ with $S_i\in\{X,Y\}$, with $n_Y$ being the number of $Y$ operator in $S$, the cat state has the following feature 
\eq{\label{eq: string}
\bra{\text{cat}}S \ket{\text{cat}}=\left\{
\begin{array}{cc}
+1 & \text{if }n_Y =0\mod 4,\\
-1 & \text{if }n_Y =2\mod 4,\\
0 & \text{otherwise}.\\
\end{array}
\right.}
Therefore, the decoder needs to know $n_Y$ in order to correctly determine the high-order correlation among the outcomes.
\end{enumerate}
We can see that Atlas correctly groups the observable sequences according to both aspects, providing all the necessary information for the decoder. The Boreas takes the aspect 1 into account, but groups all pure-$X/Y$ sequences within the same large cluster without clear distinction, so it cannot convey information about the aspect 2 to the decoder. This prevents Boreas from recognizing the quantum coherence of the cat state. Cygnus does not get either aspect right. Instead, it loosely groups all observable sequences based on what the first and last observables are. However, this classification seems to have little practical significance for informing the measurement outcomes.

As the information bottleneck strengthens, different clusters are forced to merge. In comparison to Atlas, Boreas choose to merge all pure-$X/Y$ sequences into a single cluster. The motivation to differentiate these sequences originally stems from the high-order correlations present in the classical shadow data, as described by \eqnref{eq: string}. However, because these high-order correlations are high-variance statistical features, they are the first to be discarded under the pressure of information bottleneck. This leads to the emergence of classicality.

\section*{Discussion}

\subsection*{Implication of Results}

In this research, we investigate the potential of generative language models for modeling classical shadows collected from randomized Pauli measurements on quantum many-body states. We specifically focus on the GHZ state, an idealized representation of Schrödinger's cat. Our findings indicate that as the size of the quantum system increases, the language model rapidly loses its grasp of quantum coherence. This is because quantum coherence, encoded as high-order correlations in the data, has a variance that escalates exponentially with the system size.

This phenomenon ushers in a boundary between quantum and classical realities, which we quantitatively delineate in \figref{fig: phase}. Interestingly, we discover that this boundary is not absolute, but rather influenced by the model's inherent capacity to process classical information. A more potent model can push the quantum-classical boundary towards larger system sizes. In fact, if we conduct classical shadow tomography directly based on the data, we can precisely reconstruct the full quantum state for any system size, even though the data and computational resources required for this operation also grow exponentially with system size.

Our findings suggest that our ability to process classical information may restrict our perception of the quantum essence of the universe. Despite the quantum nature of the universe, our daily experiences are predominantly classical, a perception that might stem from our limitations as classical intelligent agents.

More practically, our discoveries pose challenges to the use of deep generative models in quantum state tomography. \cite{Torlai2017N1703.05334, Torlai2018L1801.09684, Xu2018N1811.06654, Torlai2019I1904.08441, Neugebauer2020N2007.16185, Ahmed2021Q2008.03240, Koutny2022N2206.06736, Quek2018A1812.06693, Iouchtchenko2023N2206.15449, Carrasquilla2018R1810.10584, Carrasquilla2021P1912.11052, Cha2022A2006.12469} It's crucial to acknowledge that a model might not necessarily capture all statistical features in the data through training, particularly those high-order correlations. \cite{Goldt2020T2006.14709, Ingrosso2022D2202.00565, Refinetti2022N2211.11567} As a result, for larger quantum systems, generative models might struggle to fully reconstruct quantum coherence and entanglement. This makes it difficult to avoid a certain degree of decoherence in the reconstruction results.

\subsection*{Related Works}

Our research aligns and intersects with existing work in the following domains:
\begin{itemize}
\item Emergent Classicality: Some studies \cite{Brandao2015G1310.8640, Foti2019W1810.10261, Qi2021E2001.01507, Coppo2022T2203.13587} have analyzed emergent classicality from the perspective of partial observation. When a portion of a quantum system (the quantum Markov blanket) is excluded from observation during the data acquisition phase, any locally accessible information about the remaining observable subsystem will appear classical. In contrast, our work illustrates the emergence of classicality from an information bottleneck in the classical post-processing phase. This emergence occurs even when every qubit of the quantum system is observed, suggesting that a lossy compression encoding of the observable sequence can also lead to the emergence of classicality.

\item Machine-Learning Quantum State Tomography (MLQST): The objective of MLQST is to employ machine learning models to facilitate an efficient representation of quantum states. The combination of the generative language model with classical shadow tomography in this work can be viewed as a strategy for MLQST. Our approach does not directly use a neural network to model the quantum state itself, instead, we employ a generative model to learn the probability distribution of the measurement outcomes under random measurements of the quantum state. This approach diverges from many neural-network-based MLQST methods \cite{Torlai2017N1703.05334, Torlai2018L1801.09684, Xu2018N1811.06654, Torlai2019I1904.08441, Neugebauer2020N2007.16185, Ahmed2021Q2008.03240, Koutny2022N2206.06736} that rely on direct modeling of the quantum state. Additionally, in terms of the technical approach to quantum state reconstruction from randomized measurements, we follow the classical shadow reconstruction rather than positive operator-valued measure (POVM) inversion\cite{Carrasquilla2018R1810.10584, Carrasquilla2021P1912.11052, Cha2022A2006.12469}. This choice grants us more flexibility in the selection of the measurement basis.

\item Classical Shadows and Machine Learning: Classical shadow tomography provides an effective interface for the mutual conversion between quantum states and classical data. Consequently, it is perceived as a crucial integration point between quantum information and machine learning. Numerous studies \cite{Li2020V2012.08288, Huang2021P2011.01938, Huang2021I2101.02464, Huang2021P2106.12627, Huang2022Q2112.00778, Van-Kirk2022H2212.06084, Wei2023N2305.01078, Jerbi2023S2306.00061} have showcased the superiority of machine learning algorithms in classifying or interpolating quantum states based on classical shadow data, with the majority of these studies concentrating on supervised learning. Our research delves into the realm of unsupervised generative modeling of classical shadows, demonstrating the feasibility of driving representation learning of quantum observables through classical shadow data.
\end{itemize}

\subsection*{Future Directions}

Many advances in deep learning are based on representation learning, which transforms complex data like images and language into a more manageable latent space. Extending this idea to quantum information, we aim to let artificial intelligence comprehend the ``language'' of quantum states and quantum operators through representation learning \cite{Iten2020D1807.10300, Poulsen-Nautrup2020O2001.00593, Frohnert2023E2306.05694}. However, this process requires a vast amount of data.

Our research showcases the representation learning of quantum observables, as illustrated in \figref{fig: starchart}. We demonstrate that randomized measurement serves as a potent data source, capable of providing a large amount of unlabeled data for generative models. Such data can now be conveniently acquired on Noisy Intermediate-Scale Quantum (NISQ)\cite{Preskill2018Q1801.00862} devices. Utilizing these data to train dedicated language models could provide foundational models for quantum many-body physics. The learned latent representations can also support numerous downstream applications, contributing to our ultimate goal of building AI quantum physicists.

There are a few future directions to explore. First, unconstraint generative modeling of classical shadows may produce non-physical states (indefinite density matrices). The question is, how can we restrict the probability space to the physical subspace? One possible solution could be adversarial learning, which introduces a discriminator to keep the generator from breaking the positivity bound of the reconstructed state. Also, another pressing issue is to go beyond single-qubit Pauli measurements to gain advantages from quantum entanglement. Recent advancements in shallow-circuit classical shadow tomography\cite{Hu2022H2102.10132, Hu2023C2107.04817, Akhtar2023S2209.02093, Bertoni2022S2209.12924, Ippoliti2023O2212.11963} have made promising strides. It allows the extension of random measurements to commuting multi-qubit observables, thereby improving measurement efficiency. However, translating these classical shadow data into a format suitable for language models, and integrating them with generative models, remains a future research direction.

\bibliography{reference}

\begin{thebibliography}{10}
\urlstyle{rm}
\expandafter\ifx\csname url\endcsname\relax
  \def\url#1{\texttt{#1}}\fi
\expandafter\ifx\csname urlprefix\endcsname\relax\def\urlprefix{URL }\fi
\expandafter\ifx\csname doiprefix\endcsname\relax\def\doiprefix{DOI: }\fi
\providecommand{\bibinfo}[2]{#2}
\providecommand{\eprint}[2][]{\url{#2}}

\bibitem{Leggett2002T}
\bibinfo{author}{{Leggett}, A.~J.}
\newblock \bibinfo{journal}{\bibinfo{title}{{TOPICAL REVIEW: Testing the limits
  of quantum mechanics: motivation, state of play, prospects}}}.
\newblock {\emph{\JournalTitle{Journal of Physics Condensed Matter}}}
  \textbf{\bibinfo{volume}{14}}, \bibinfo{pages}{R415--R451},
  \doiprefix\url{10.1088/0953-8984/14/15/201} (\bibinfo{year}{2002}).

\bibitem{Schlosshauer2014T1404.2635}
\bibinfo{author}{{Schlosshauer}, M.}
\newblock \bibinfo{journal}{\bibinfo{title}{{The quantum-to-classical
  transition and decoherence}}}.
\newblock {\emph{\JournalTitle{arXiv e-prints}}}
  \bibinfo{pages}{arXiv:1404.2635}, \doiprefix\url{10.48550/arXiv.1404.2635}
  (\bibinfo{year}{2014}).
\newblock \eprint{1404.2635}.

\bibitem{Schrodinger1935D}
\bibinfo{author}{Schr{\"o}dinger, E.}
\newblock \bibinfo{journal}{\bibinfo{title}{Die gegenw{\"a}rtige situation in
  der quantenmechanik}}.
\newblock {\emph{\JournalTitle{Naturwissenschaften}}}
  \textbf{\bibinfo{volume}{23}}, \bibinfo{pages}{807--812},
  \doiprefix\url{10.1007/BF01491891} (\bibinfo{year}{1935}).

\bibitem{Zeh1970O}
\bibinfo{author}{Zeh, H.~D.}
\newblock \bibinfo{journal}{\bibinfo{title}{On the interpretation of
  measurement in quantum theory}}.
\newblock {\emph{\JournalTitle{Foundations of Physics}}}
  \textbf{\bibinfo{volume}{1}}, \bibinfo{pages}{69--76},
  \doiprefix\url{10.1007/BF00708656} (\bibinfo{year}{1970}).

\bibitem{Joos1985Emergence}
\bibinfo{author}{Joos, E.} \& \bibinfo{author}{Zeh, H.~D.}
\newblock \bibinfo{journal}{\bibinfo{title}{The emergence of classical
  properties through interaction with the environment}}.
\newblock {\emph{\JournalTitle{Zeitschrift f{\"u}r Physik B Condensed Matter}}}
  \textbf{\bibinfo{volume}{59}}, \bibinfo{pages}{223--243},
  \doiprefix\url{10.1007/BF01725541} (\bibinfo{year}{1985}).

\bibitem{Schlosshauer2004Dquant-ph/0312059}
\bibinfo{author}{{Schlosshauer}, M.}
\newblock \bibinfo{journal}{\bibinfo{title}{{Decoherence, the measurement
  problem, and interpretations of quantum mechanics}}}.
\newblock {\emph{\JournalTitle{Reviews of Modern Physics}}}
  \textbf{\bibinfo{volume}{76}}, \bibinfo{pages}{1267--1305},
  \doiprefix\url{10.1103/RevModPhys.76.1267} (\bibinfo{year}{2004}).
\newblock \eprint{quant-ph/0312059}.

\bibitem{Zurek1981Pointer}
\bibinfo{author}{Zurek, W.~H.}
\newblock \bibinfo{journal}{\bibinfo{title}{Pointer basis of quantum apparatus:
  Into what mixture does the wave packet collapse?}}
\newblock {\emph{\JournalTitle{Physical Review D}}}
  \textbf{\bibinfo{volume}{24}}, \bibinfo{pages}{1516--1525},
  \doiprefix\url{10.1103/PhysRevD.24.1516} (\bibinfo{year}{1981}).

\bibitem{Zurek1982Environment-induced}
\bibinfo{author}{Zurek, W.~H.}
\newblock \bibinfo{journal}{\bibinfo{title}{Environment-induced superselection
  rules}}.
\newblock {\emph{\JournalTitle{Physical Review D}}}
  \textbf{\bibinfo{volume}{26}}, \bibinfo{pages}{1862--1880},
  \doiprefix\url{10.1103/PhysRevD.26.1862} (\bibinfo{year}{1982}).

\bibitem{Zurek1993Preferred}
\bibinfo{author}{Zurek, W.~H.}
\newblock \bibinfo{journal}{\bibinfo{title}{{Preferred States, Predictability,
  Classicality and the Environment-Induced Decoherence}}}.
\newblock {\emph{\JournalTitle{Progress of Theoretical Physics}}}
  \textbf{\bibinfo{volume}{89}}, \bibinfo{pages}{281--312},
  \doiprefix\url{10.1143/ptp/89.2.281} (\bibinfo{year}{1993}).
\newblock
  \eprint{https://academic.oup.com/ptp/article-pdf/89/2/281/5226677/89-2-281.pdf}.

\bibitem{Zurek1998Decoherence}
\bibinfo{author}{{Zurek}, W.~H.}
\newblock \bibinfo{journal}{\bibinfo{title}{{Decoherence, einselection and the
  existential interpretation (the rough guide)}}}.
\newblock {\emph{\JournalTitle{Philosophical Transactions of the Royal Society
  of London Series A}}} \textbf{\bibinfo{volume}{356}}, \bibinfo{pages}{1793},
  \doiprefix\url{10.1098/rsta.1998.0250} (\bibinfo{year}{1998}).
\newblock \eprint{quant-ph/9805065}.

\bibitem{Zurek2009Quantum}
\bibinfo{author}{{Zurek}, W.~H.}
\newblock \bibinfo{journal}{\bibinfo{title}{{Quantum Darwinism}}}.
\newblock {\emph{\JournalTitle{Nature Physics}}} \textbf{\bibinfo{volume}{5}},
  \bibinfo{pages}{181--188}, \doiprefix\url{10.1038/nphys1202}
  (\bibinfo{year}{2009}).
\newblock \eprint{0903.5082}.

\bibitem{Everett1957R}
\bibinfo{author}{Everett, H.}
\newblock \bibinfo{journal}{\bibinfo{title}{"relative state" formulation of
  quantum mechanics}}.
\newblock {\emph{\JournalTitle{Rev. Mod. Phys.}}}
  \textbf{\bibinfo{volume}{29}}, \bibinfo{pages}{454--462},
  \doiprefix\url{10.1103/RevModPhys.29.454} (\bibinfo{year}{1957}).

\bibitem{Everett2015T}
\bibinfo{author}{Everett, H.}
\newblock \bibinfo{title}{The theory of the universal wave function}.
\newblock In \bibinfo{editor}{DeWitt, B.~S.} \& \bibinfo{editor}{Graham, N.}
  (eds.) \emph{\bibinfo{booktitle}{The many-worlds interpretation of quantum
  mechanics}}, vol.~\bibinfo{volume}{61}, chap.~\bibinfo{chapter}{1},
  \bibinfo{pages}{1} (\bibinfo{publisher}{Princeton University Press},
  \bibinfo{year}{2015}).

\bibitem{Tegmark1998Tquant-ph/9709032}
\bibinfo{author}{{Tegmark}, M.}
\newblock \bibinfo{journal}{\bibinfo{title}{{The Interpretation of Quantum
  Mechanics: Many Worlds or Many Words?}}}
\newblock {\emph{\JournalTitle{Fortschritte der Physik}}}
  \textbf{\bibinfo{volume}{46}}, \bibinfo{pages}{855--862},
  \doiprefix\url{10.1002/(SICI)1521-3978(199811)46:6/8<855::AID-PROP855>3.0.CO;2-Q}
  (\bibinfo{year}{1998}).
\newblock \eprint{quant-ph/9709032}.

\bibitem{Ghirardi1986U}
\bibinfo{author}{Ghirardi, G.~C.}, \bibinfo{author}{Rimini, A.} \&
  \bibinfo{author}{Weber, T.}
\newblock \bibinfo{journal}{\bibinfo{title}{Unified dynamics for microscopic
  and macroscopic systems}}.
\newblock {\emph{\JournalTitle{Physical Review D}}}
  \textbf{\bibinfo{volume}{34}}, \bibinfo{pages}{470--491},
  \doiprefix\url{10.1103/PhysRevD.34.470} (\bibinfo{year}{1986}).

\bibitem{Ghirardi1990M}
\bibinfo{author}{Ghirardi, G.~C.}, \bibinfo{author}{Pearle, P.} \&
  \bibinfo{author}{Rimini, A.}
\newblock \bibinfo{journal}{\bibinfo{title}{Markov processes in hilbert space
  and continuous spontaneous localization of systems of identical particles}}.
\newblock {\emph{\JournalTitle{Physical Review A}}}
  \textbf{\bibinfo{volume}{42}}, \bibinfo{pages}{78--89},
  \doiprefix\url{10.1103/PhysRevA.42.78} (\bibinfo{year}{1990}).

\bibitem{Bassi2013Models}
\bibinfo{author}{{Bassi}, A.}, \bibinfo{author}{{Lochan}, K.},
  \bibinfo{author}{{Satin}, S.}, \bibinfo{author}{{Singh}, T.~P.} \&
  \bibinfo{author}{{Ulbricht}, H.}
\newblock \bibinfo{journal}{\bibinfo{title}{{Models of wave-function collapse,
  underlying theories, and experimental tests}}}.
\newblock {\emph{\JournalTitle{Reviews of Modern Physics}}}
  \textbf{\bibinfo{volume}{85}}, \bibinfo{pages}{471--527},
  \doiprefix\url{10.1103/RevModPhys.85.471} (\bibinfo{year}{2013}).
\newblock \eprint{1204.4325}.

\bibitem{Home1992E}
\bibinfo{author}{Home, D.} \& \bibinfo{author}{Whitaker, M.}
\newblock \bibinfo{journal}{\bibinfo{title}{Ensemble interpretations of quantum
  mechanics. a modern perspective}}.
\newblock {\emph{\JournalTitle{Physics Reports}}}
  \textbf{\bibinfo{volume}{210}}, \bibinfo{pages}{223--317},
  \doiprefix\url{https://doi.org/10.1016/0370-1573(92)90088-H}
  (\bibinfo{year}{1992}).

\bibitem{Fuchs2001Qquant-ph/0106166}
\bibinfo{author}{{Fuchs}, C.~A.}
\newblock \bibinfo{journal}{\bibinfo{title}{{Quantum Foundations in the Light
  of Quantum Information}}}.
\newblock {\emph{\JournalTitle{arXiv e-prints}}}
  \bibinfo{pages}{quant--ph/0106166},
  \doiprefix\url{10.48550/arXiv.quant-ph/0106166} (\bibinfo{year}{2001}).
\newblock \eprint{quant-ph/0106166}.

\bibitem{Caves2002Qquant-ph/0106133}
\bibinfo{author}{{Caves}, C.~M.}, \bibinfo{author}{{Fuchs}, C.~A.} \&
  \bibinfo{author}{{Schack}, R.}
\newblock \bibinfo{journal}{\bibinfo{title}{{Quantum probabilities as Bayesian
  probabilities}}}.
\newblock {\emph{\JournalTitle{Physical Review A}}}
  \textbf{\bibinfo{volume}{65}}, \bibinfo{pages}{022305},
  \doiprefix\url{10.1103/PhysRevA.65.022305} (\bibinfo{year}{2002}).
\newblock \eprint{quant-ph/0106133}.

\bibitem{Fuchs2002Qquant-ph/0205039}
\bibinfo{author}{{Fuchs}, C.~A.}
\newblock \bibinfo{journal}{\bibinfo{title}{{Quantum Mechanics as Quantum
  Information (and only a little more)}}}.
\newblock {\emph{\JournalTitle{arXiv e-prints}}}
  \bibinfo{pages}{quant--ph/0205039},
  \doiprefix\url{10.48550/arXiv.quant-ph/0205039} (\bibinfo{year}{2002}).
\newblock \eprint{quant-ph/0205039}.

\bibitem{Caves2002Cquant-ph/0206110}
\bibinfo{author}{{Caves}, C.~M.}, \bibinfo{author}{{Fuchs}, C.~A.} \&
  \bibinfo{author}{{Schack}, R.}
\newblock \bibinfo{journal}{\bibinfo{title}{{Conditions for compatibility of
  quantum-state assignments}}}.
\newblock {\emph{\JournalTitle{Physical Review A}}}
  \textbf{\bibinfo{volume}{66}}, \bibinfo{pages}{062111},
  \doiprefix\url{10.1103/PhysRevA.66.062111} (\bibinfo{year}{2002}).
\newblock \eprint{quant-ph/0206110}.

\bibitem{Fuchs2004U}
\bibinfo{author}{{Fuchs}, C.~A.} \& \bibinfo{author}{{Schack}, R.}
\newblock \bibinfo{title}{{Unknown Quantum States and Operations,a Bayesian
  View}}.
\newblock In \bibinfo{editor}{{Paris}, M. G.~A.} \&
  \bibinfo{editor}{{{\v{R}}eh{\'a}{\v{c}}ek}, J.} (eds.)
  \emph{\bibinfo{booktitle}{Quantum State Estimation}}, vol.
  \bibinfo{volume}{649}, \bibinfo{pages}{147--187},
  \doiprefix\url{10.1007/978-3-540-44481-7_5} (\bibinfo{publisher}{Springer},
  \bibinfo{year}{2004}).

\bibitem{Fuchs2013Q1301.3274}
\bibinfo{author}{{Fuchs}, C.~A.} \& \bibinfo{author}{{Schack}, R.}
\newblock \bibinfo{journal}{\bibinfo{title}{{Quantum-Bayesian coherence}}}.
\newblock {\emph{\JournalTitle{Reviews of Modern Physics}}}
  \textbf{\bibinfo{volume}{85}}, \bibinfo{pages}{1693--1715},
  \doiprefix\url{10.1103/RevModPhys.85.1693} (\bibinfo{year}{2013}).
\newblock \eprint{1301.3274}.

\bibitem{Mermin2014P}
\bibinfo{author}{Mermin, N.~D.}
\newblock \bibinfo{journal}{\bibinfo{title}{Physics: Qbism puts the scientist
  back into science}}.
\newblock {\emph{\JournalTitle{Nature}}} \textbf{\bibinfo{volume}{507}},
  \bibinfo{pages}{421--423}, \doiprefix\url{10.1038/507421a}
  (\bibinfo{year}{2014}).

\bibitem{Brandao2015G1310.8640}
\bibinfo{author}{{Brand{\~a}o}, F. G.~S.~L.}, \bibinfo{author}{{Piani}, M.} \&
  \bibinfo{author}{{Horodecki}, P.}
\newblock \bibinfo{journal}{\bibinfo{title}{{Generic emergence of classical
  features in quantum Darwinism}}}.
\newblock {\emph{\JournalTitle{Nature Communications}}}
  \textbf{\bibinfo{volume}{6}}, \bibinfo{pages}{7908},
  \doiprefix\url{10.1038/ncomms8908} (\bibinfo{year}{2015}).
\newblock \eprint{1310.8640}.

\bibitem{Foti2019W1810.10261}
\bibinfo{author}{{Foti}, C.}, \bibinfo{author}{{Heinosaari}, T.},
  \bibinfo{author}{{Maniscalco}, S.} \& \bibinfo{author}{{Verrucchi}, P.}
\newblock \bibinfo{journal}{\bibinfo{title}{{Whenever a quantum environment
  emerges as a classical system, it behaves like a measuring apparatus}}}.
\newblock {\emph{\JournalTitle{Quantum}}} \textbf{\bibinfo{volume}{3}},
  \bibinfo{pages}{179}, \doiprefix\url{10.22331/q-2019-08-26-179}
  (\bibinfo{year}{2019}).
\newblock \eprint{1810.10261}.

\bibitem{Qi2021E2001.01507}
\bibinfo{author}{{Qi}, X.-L.} \& \bibinfo{author}{{Ranard}, D.}
\newblock \bibinfo{journal}{\bibinfo{title}{{Emergent classicality in general
  multipartite states and channels}}}.
\newblock {\emph{\JournalTitle{Quantum}}} \textbf{\bibinfo{volume}{5}},
  \bibinfo{pages}{555}, \doiprefix\url{10.22331/q-2021-09-28-555}
  (\bibinfo{year}{2021}).
\newblock \eprint{2001.01507}.

\bibitem{Coppo2022T2203.13587}
\bibinfo{author}{{Coppo}, A.}, \bibinfo{author}{{Pranzini}, N.} \&
  \bibinfo{author}{{Verrucchi}, P.}
\newblock \bibinfo{journal}{\bibinfo{title}{{Threshold size for the emergence
  of classical-like behavior}}}.
\newblock {\emph{\JournalTitle{Physical Review A}}}
  \textbf{\bibinfo{volume}{106}}, \bibinfo{pages}{042208},
  \doiprefix\url{10.1103/PhysRevA.106.042208} (\bibinfo{year}{2022}).
\newblock \eprint{2203.13587}.

\bibitem{Paris2004Q}
\bibinfo{author}{Paris, M.} \& \bibinfo{author}{Rehacek, J.}
\newblock \emph{\bibinfo{title}{Quantum state estimation}}, vol.
  \bibinfo{volume}{649} (\bibinfo{publisher}{Springer Science \& Business
  Media}, \bibinfo{year}{2004}).

\bibitem{Cramer2010E1101.4366}
\bibinfo{author}{{Cramer}, M.} \emph{et~al.}
\newblock \bibinfo{journal}{\bibinfo{title}{{Efficient quantum state
  tomography}}}.
\newblock {\emph{\JournalTitle{Nature Communications}}}
  \textbf{\bibinfo{volume}{1}}, \bibinfo{pages}{149},
  \doiprefix\url{10.1038/ncomms1147} (\bibinfo{year}{2010}).
\newblock \eprint{1101.4366}.

\bibitem{Flammia2012Q1205.2300}
\bibinfo{author}{{Flammia}, S.~T.}, \bibinfo{author}{{Gross}, D.},
  \bibinfo{author}{{Liu}, Y.-K.} \& \bibinfo{author}{{Eisert}, J.}
\newblock \bibinfo{journal}{\bibinfo{title}{{Quantum tomography via compressed
  sensing: error bounds, sample complexity and efficient estimators}}}.
\newblock {\emph{\JournalTitle{New Journal of Physics}}}
  \textbf{\bibinfo{volume}{14}}, \bibinfo{pages}{095022},
  \doiprefix\url{10.1088/1367-2630/14/9/095022} (\bibinfo{year}{2012}).
\newblock \eprint{1205.2300}.

\bibitem{ODonnell2015E1508.01907}
\bibinfo{author}{{O'Donnell}, R.} \& \bibinfo{author}{{Wright}, J.}
\newblock \bibinfo{journal}{\bibinfo{title}{{Efficient quantum tomography}}}.
\newblock {\emph{\JournalTitle{arXiv e-prints}}}
  \bibinfo{pages}{arXiv:1508.01907}, \doiprefix\url{10.48550/arXiv.1508.01907}
  (\bibinfo{year}{2015}).
\newblock \eprint{1508.01907}.

\bibitem{Haah2015S1508.01797}
\bibinfo{author}{{Haah}, J.}, \bibinfo{author}{{Harrow}, A.~W.},
  \bibinfo{author}{{Ji}, Z.}, \bibinfo{author}{{Wu}, X.} \&
  \bibinfo{author}{{Yu}, N.}
\newblock \bibinfo{journal}{\bibinfo{title}{{Sample-optimal tomography of
  quantum states}}}.
\newblock {\emph{\JournalTitle{arXiv e-prints}}}
  \bibinfo{pages}{arXiv:1508.01797}, \doiprefix\url{10.48550/arXiv.1508.01797}
  (\bibinfo{year}{2015}).
\newblock \eprint{1508.01797}.

\bibitem{Lanyon2017E1612.08000}
\bibinfo{author}{{Lanyon}, B.~P.} \emph{et~al.}
\newblock \bibinfo{journal}{\bibinfo{title}{{Efficient tomography of a quantum
  many-body system}}}.
\newblock {\emph{\JournalTitle{Nature Physics}}} \textbf{\bibinfo{volume}{13}},
  \bibinfo{pages}{1158--1162}, \doiprefix\url{10.1038/nphys4244}
  (\bibinfo{year}{2017}).
\newblock \eprint{1612.08000}.

\bibitem{Brandao2017Q1710.02581}
\bibinfo{author}{{Brand{\~a}o}, F. G.~S.~L.} \emph{et~al.}
\newblock \bibinfo{journal}{\bibinfo{title}{{Quantum SDP Solvers: Large
  Speed-ups, Optimality, and Applications to Quantum Learning}}}.
\newblock {\emph{\JournalTitle{arXiv e-prints}}}
  \bibinfo{pages}{arXiv:1710.02581} (\bibinfo{year}{2017}).
\newblock \eprint{1710.02581}.

\bibitem{Aaronson2017S1711.01053}
\bibinfo{author}{{Aaronson}, S.}
\newblock \bibinfo{journal}{\bibinfo{title}{{Shadow Tomography of Quantum
  States}}}.
\newblock {\emph{\JournalTitle{arXiv e-prints}}}
  \bibinfo{pages}{arXiv:1711.01053} (\bibinfo{year}{2017}).
\newblock \eprint{1711.01053}.

\bibitem{Wang2017S1712.03213}
\bibinfo{author}{{Wang}, J.} \emph{et~al.}
\newblock \bibinfo{journal}{\bibinfo{title}{{Scalable Quantum Tomography with
  Fidelity Estimation}}}.
\newblock {\emph{\JournalTitle{arXiv e-prints}}}
  \bibinfo{pages}{arXiv:1712.03213} (\bibinfo{year}{2017}).
\newblock \eprint{1712.03213}.

\bibitem{Aaronson2019G1904.08747}
\bibinfo{author}{{Aaronson}, S.} \& \bibinfo{author}{{Rothblum}, G.~N.}
\newblock \bibinfo{journal}{\bibinfo{title}{{Gentle Measurement of Quantum
  States and Differential Privacy}}}.
\newblock {\emph{\JournalTitle{arXiv e-prints}}}
  \bibinfo{pages}{arXiv:1904.08747} (\bibinfo{year}{2019}).
\newblock \eprint{1904.08747}.

\bibitem{Ohliger2013E1204.5735}
\bibinfo{author}{{Ohliger}, M.}, \bibinfo{author}{{Nesme}, V.} \&
  \bibinfo{author}{{Eisert}, J.}
\newblock \bibinfo{journal}{\bibinfo{title}{{Efficient and feasible state
  tomography of quantum many-body systems}}}.
\newblock {\emph{\JournalTitle{New Journal of Physics}}}
  \textbf{\bibinfo{volume}{15}}, \bibinfo{pages}{015024},
  \doiprefix\url{10.1088/1367-2630/15/1/015024} (\bibinfo{year}{2013}).
\newblock \eprint{1204.5735}.

\bibitem{Guta2018F1809.11162}
\bibinfo{author}{{Guta}, M.}, \bibinfo{author}{{Kahn}, J.},
  \bibinfo{author}{{Kueng}, R.} \& \bibinfo{author}{{Tropp}, J.~A.}
\newblock \bibinfo{journal}{\bibinfo{title}{{Fast state tomography with optimal
  error bounds}}}.
\newblock {\emph{\JournalTitle{Journal of Physics A: Mathematical and
  Theoretical}}} \textbf{\bibinfo{volume}{53}}, \bibinfo{pages}{204001},
  \doiprefix\url{10.1088/1751-8121/ab8111} (\bibinfo{year}{2020}).
\newblock \eprint{1809.11162}.

\bibitem{Huang2020P2002.08953}
\bibinfo{author}{{Huang}, H.-Y.}, \bibinfo{author}{{Kueng}, R.} \&
  \bibinfo{author}{{Preskill}, J.}
\newblock \bibinfo{journal}{\bibinfo{title}{{Predicting many properties of a
  quantum system from very few measurements}}}.
\newblock {\emph{\JournalTitle{Nature Physics}}} \textbf{\bibinfo{volume}{16}},
  \bibinfo{pages}{1050--1057}, \doiprefix\url{10.1038/s41567-020-0932-7}
  (\bibinfo{year}{2020}).
\newblock \eprint{2002.08953}.

\bibitem{Radford2018I}
\bibinfo{author}{Radford, A.}, \bibinfo{author}{Narasimhan, K.},
  \bibinfo{author}{Salimans, T.} \& \bibinfo{author}{Sutskever, I.}
\newblock \bibinfo{title}{Improving language understanding by generative
  pre-training}.
\newblock \bibinfo{type}{Tech. Rep.}, \bibinfo{institution}{OpenAI}
  (\bibinfo{year}{2018}).

\bibitem{Tishby2000Tphysics/0004057}
\bibinfo{author}{{Tishby}, N.}, \bibinfo{author}{{Pereira}, F.~C.} \&
  \bibinfo{author}{{Bialek}, W.}
\newblock \bibinfo{journal}{\bibinfo{title}{{The information bottleneck
  method}}}.
\newblock {\emph{\JournalTitle{arXiv e-prints}}}
  \bibinfo{pages}{physics/0004057},
  \doiprefix\url{10.48550/arXiv.physics/0004057} (\bibinfo{year}{2000}).
\newblock \eprint{physics/0004057}.

\bibitem{Tishby2015D1503.02406}
\bibinfo{author}{{Tishby}, N.} \& \bibinfo{author}{{Zaslavsky}, N.}
\newblock \bibinfo{journal}{\bibinfo{title}{{Deep Learning and the Information
  Bottleneck Principle}}}.
\newblock {\emph{\JournalTitle{arXiv e-prints}}}
  \bibinfo{pages}{arXiv:1503.02406}, \doiprefix\url{10.48550/arXiv.1503.02406}
  (\bibinfo{year}{2015}).
\newblock \eprint{1503.02406}.

\bibitem{Fisher2015Q1508.05929}
\bibinfo{author}{{Fisher}, M. P.~A.}
\newblock \bibinfo{journal}{\bibinfo{title}{{Quantum cognition: The possibility
  of processing with nuclear spins in the brain}}}.
\newblock {\emph{\JournalTitle{Annals of Physics}}}
  \textbf{\bibinfo{volume}{362}}, \bibinfo{pages}{593--602},
  \doiprefix\url{10.1016/j.aop.2015.08.020} (\bibinfo{year}{2015}).
\newblock \eprint{1508.05929}.

\bibitem{Greenberger2007G0712.0921}
\bibinfo{author}{{Greenberger}, D.~M.}, \bibinfo{author}{{Horne}, M.~A.} \&
  \bibinfo{author}{{Zeilinger}, A.}
\newblock \bibinfo{journal}{\bibinfo{title}{{Going Beyond Bell's Theorem}}}.
\newblock {\emph{\JournalTitle{arXiv e-prints}}}
  \bibinfo{pages}{arXiv:0712.0921}, \doiprefix\url{10.48550/arXiv.0712.0921}
  (\bibinfo{year}{2007}).
\newblock \eprint{0712.0921}.

\bibitem{Raimond2001M}
\bibinfo{author}{Raimond, J.~M.}, \bibinfo{author}{Brune, M.} \&
  \bibinfo{author}{Haroche, S.}
\newblock \bibinfo{journal}{\bibinfo{title}{Manipulating quantum entanglement
  with atoms and photons in a cavity}}.
\newblock {\emph{\JournalTitle{Rev. Mod. Phys.}}}
  \textbf{\bibinfo{volume}{73}}, \bibinfo{pages}{565--582},
  \doiprefix\url{10.1103/RevModPhys.73.565} (\bibinfo{year}{2001}).

\bibitem{Nielsen2000Q}
\bibinfo{author}{Nielsen, M.~A.} \& \bibinfo{author}{Chuang, I.~L.}
\newblock \emph{\bibinfo{title}{Quantum computation and quantum information}}
  (\bibinfo{publisher}{Cambridge University Press, Cambridge},
  \bibinfo{year}{2000}).

\bibitem{Li2020V2012.08288}
\bibinfo{author}{{Li}, G.}, \bibinfo{author}{{Song}, Z.} \&
  \bibinfo{author}{{Wang}, X.}
\newblock \bibinfo{journal}{\bibinfo{title}{{VSQL: Variational Shadow Quantum
  Learning for Classification}}}.
\newblock {\emph{\JournalTitle{arXiv e-prints}}}
  \bibinfo{pages}{arXiv:2012.08288}, \doiprefix\url{10.48550/arXiv.2012.08288}
  (\bibinfo{year}{2020}).
\newblock \eprint{2012.08288}.

\bibitem{Huang2021P2011.01938}
\bibinfo{author}{{Huang}, H.-Y.} \emph{et~al.}
\newblock \bibinfo{journal}{\bibinfo{title}{{Power of data in quantum machine
  learning}}}.
\newblock {\emph{\JournalTitle{Nature Communications}}}
  \textbf{\bibinfo{volume}{12}}, \bibinfo{pages}{2631},
  \doiprefix\url{10.1038/s41467-021-22539-9} (\bibinfo{year}{2021}).
\newblock \eprint{2011.01938}.

\bibitem{Huang2021I2101.02464}
\bibinfo{author}{{Huang}, H.-Y.}, \bibinfo{author}{{Kueng}, R.} \&
  \bibinfo{author}{{Preskill}, J.}
\newblock \bibinfo{journal}{\bibinfo{title}{{Information-Theoretic Bounds on
  Quantum Advantage in Machine Learning}}}.
\newblock {\emph{\JournalTitle{Physical Review Letters}}}
  \textbf{\bibinfo{volume}{126}}, \bibinfo{pages}{190505},
  \doiprefix\url{10.1103/PhysRevLett.126.190505} (\bibinfo{year}{2021}).
\newblock \eprint{2101.02464}.

\bibitem{Huang2021P2106.12627}
\bibinfo{author}{{Huang}, H.-Y.}, \bibinfo{author}{{Kueng}, R.},
  \bibinfo{author}{{Torlai}, G.}, \bibinfo{author}{{Albert}, V.~V.} \&
  \bibinfo{author}{{Preskill}, J.}
\newblock \bibinfo{journal}{\bibinfo{title}{{Provably efficient machine
  learning for quantum many-body problems}}}.
\newblock {\emph{\JournalTitle{arXiv e-prints}}}
  \bibinfo{pages}{arXiv:2106.12627}, \doiprefix\url{10.48550/arXiv.2106.12627}
  (\bibinfo{year}{2021}).
\newblock \eprint{2106.12627}.

\bibitem{Huang2022Q2112.00778}
\bibinfo{author}{{Huang}, H.-Y.} \emph{et~al.}
\newblock \bibinfo{journal}{\bibinfo{title}{{Quantum advantage in learning from
  experiments}}}.
\newblock {\emph{\JournalTitle{Science}}} \textbf{\bibinfo{volume}{376}},
  \bibinfo{pages}{1182--1186}, \doiprefix\url{10.1126/science.abn7293}
  (\bibinfo{year}{2022}).
\newblock \eprint{2112.00778}.

\bibitem{Van-Kirk2022H2212.06084}
\bibinfo{author}{{Van Kirk}, K.}, \bibinfo{author}{{Cotler}, J.},
  \bibinfo{author}{{Huang}, H.-Y.} \& \bibinfo{author}{{Lukin}, M.~D.}
\newblock \bibinfo{journal}{\bibinfo{title}{{Hardware-efficient learning of
  quantum many-body states}}}.
\newblock {\emph{\JournalTitle{arXiv e-prints}}}
  \bibinfo{pages}{arXiv:2212.06084}, \doiprefix\url{10.48550/arXiv.2212.06084}
  (\bibinfo{year}{2022}).
\newblock \eprint{2212.06084}.

\bibitem{Wei2023N2305.01078}
\bibinfo{author}{{Wei}, V.}, \bibinfo{author}{{Coish}, W.~A.},
  \bibinfo{author}{{Ronagh}, P.} \& \bibinfo{author}{{Muschik}, C.~A.}
\newblock \bibinfo{journal}{\bibinfo{title}{{Neural-Shadow Quantum State
  Tomography}}}.
\newblock {\emph{\JournalTitle{arXiv e-prints}}}
  \bibinfo{pages}{arXiv:2305.01078}, \doiprefix\url{10.48550/arXiv.2305.01078}
  (\bibinfo{year}{2023}).
\newblock \eprint{2305.01078}.

\bibitem{Jerbi2023S2306.00061}
\bibinfo{author}{{Jerbi}, S.}, \bibinfo{author}{{Gyurik}, C.},
  \bibinfo{author}{{Marshall}, S.~C.}, \bibinfo{author}{{Molteni}, R.} \&
  \bibinfo{author}{{Dunjko}, V.}
\newblock \bibinfo{journal}{\bibinfo{title}{{Shadows of quantum machine
  learning}}}.
\newblock {\emph{\JournalTitle{arXiv e-prints}}}
  \bibinfo{pages}{arXiv:2306.00061}, \doiprefix\url{10.48550/arXiv.2306.00061}
  (\bibinfo{year}{2023}).
\newblock \eprint{2306.00061}.

\bibitem{Vaswani2017A1706.03762}
\bibinfo{author}{{Vaswani}, A.} \emph{et~al.}
\newblock \bibinfo{journal}{\bibinfo{title}{{Attention Is All You Need}}}.
\newblock {\emph{\JournalTitle{arXiv e-prints}}}
  \bibinfo{pages}{arXiv:1706.03762}, \doiprefix\url{10.48550/arXiv.1706.03762}
  (\bibinfo{year}{2017}).
\newblock \eprint{1706.03762}.

\bibitem{Higgins2017b}
\bibinfo{author}{Higgins, I.} \emph{et~al.}
\newblock \bibinfo{title}{beta-{VAE}: Learning basic visual concepts with a
  constrained variational framework}.
\newblock In \emph{\bibinfo{booktitle}{International Conference on Learning
  Representations}} (\bibinfo{year}{2017}).

\bibitem{Hinton2002S}
\bibinfo{author}{Hinton, G.~E.} \& \bibinfo{author}{Roweis, S.}
\newblock \bibinfo{journal}{\bibinfo{title}{Stochastic neighbor embedding}}.
\newblock {\emph{\JournalTitle{Advances in neural information processing
  systems}}} \textbf{\bibinfo{volume}{15}} (\bibinfo{year}{2002}).

\bibitem{van2008visualizing}
\bibinfo{author}{Van~der Maaten, L.} \& \bibinfo{author}{Hinton, G.}
\newblock \bibinfo{journal}{\bibinfo{title}{{Visualizing data using t-SNE}}}.
\newblock {\emph{\JournalTitle{Journal of machine learning research}}}
  \textbf{\bibinfo{volume}{9}} (\bibinfo{year}{2008}).

\bibitem{Torlai2017N1703.05334}
\bibinfo{author}{{Torlai}, G.} \emph{et~al.}
\newblock \bibinfo{journal}{\bibinfo{title}{Neural-network quantum state
  tomography}}.
\newblock {\emph{\JournalTitle{Nature Physics}}} \textbf{\bibinfo{volume}{14}},
  \bibinfo{pages}{447--450}, \doiprefix\url{10.1038/s41567-018-0048-5}
  (\bibinfo{year}{2018}).
\newblock \eprint{1703.05334}.

\bibitem{Torlai2018L1801.09684}
\bibinfo{author}{{Torlai}, G.} \& \bibinfo{author}{{Melko}, R.~G.}
\newblock \bibinfo{journal}{\bibinfo{title}{{Latent Space Purification via
  Neural Density Operators}}}.
\newblock {\emph{\JournalTitle{Physical Review Letters}}}
  \textbf{\bibinfo{volume}{120}}, \bibinfo{pages}{240503},
  \doiprefix\url{10.1103/PhysRevLett.120.240503} (\bibinfo{year}{2018}).
\newblock \eprint{1801.09684}.

\bibitem{Xu2018N1811.06654}
\bibinfo{author}{{Xu}, Q.} \& \bibinfo{author}{{Xu}, S.}
\newblock \bibinfo{journal}{\bibinfo{title}{{Neural network state estimation
  for full quantum state tomography}}}.
\newblock {\emph{\JournalTitle{arXiv e-prints}}}
  \bibinfo{pages}{arXiv:1811.06654} (\bibinfo{year}{2018}).
\newblock \eprint{1811.06654}.

\bibitem{Torlai2019I1904.08441}
\bibinfo{author}{{Torlai}, G.} \emph{et~al.}
\newblock \bibinfo{journal}{\bibinfo{title}{{Integrating Neural Networks with a
  Quantum Simulator for State Reconstruction}}}.
\newblock {\emph{\JournalTitle{Physical Review Letters}}}
  \textbf{\bibinfo{volume}{123}}, \bibinfo{pages}{230504},
  \doiprefix\url{10.1103/PhysRevLett.123.230504} (\bibinfo{year}{2019}).
\newblock \eprint{1904.08441}.

\bibitem{Neugebauer2020N2007.16185}
\bibinfo{author}{{Neugebauer}, M.} \emph{et~al.}
\newblock \bibinfo{journal}{\bibinfo{title}{{Neural-network quantum state
  tomography in a two-qubit experiment}}}.
\newblock {\emph{\JournalTitle{Physical Review A}}}
  \textbf{\bibinfo{volume}{102}}, \bibinfo{pages}{042604},
  \doiprefix\url{10.1103/PhysRevA.102.042604} (\bibinfo{year}{2020}).
\newblock \eprint{2007.16185}.

\bibitem{Ahmed2021Q2008.03240}
\bibinfo{author}{{Ahmed}, S.}, \bibinfo{author}{{S{\'a}nchez Mu{\~n}oz}, C.},
  \bibinfo{author}{{Nori}, F.} \& \bibinfo{author}{{Kockum}, A.~F.}
\newblock \bibinfo{journal}{\bibinfo{title}{{Quantum State Tomography with
  Conditional Generative Adversarial Networks}}}.
\newblock {\emph{\JournalTitle{Physical Review Letters}}}
  \textbf{\bibinfo{volume}{127}}, \bibinfo{pages}{140502},
  \doiprefix\url{10.1103/PhysRevLett.127.140502} (\bibinfo{year}{2021}).
\newblock \eprint{2008.03240}.

\bibitem{Koutny2022N2206.06736}
\bibinfo{author}{{Koutn{\'y}}, D.}, \bibinfo{author}{{Motka}, L.},
  \bibinfo{author}{{Hradil}, Z.}, \bibinfo{author}{{{\v{R}}eh{\'a}{\v{c}}ek},
  J.} \& \bibinfo{author}{{S{\'a}nchez-Soto}, L.~L.}
\newblock \bibinfo{journal}{\bibinfo{title}{{Neural-network quantum state
  tomography}}}.
\newblock {\emph{\JournalTitle{Physical Review A}}}
  \textbf{\bibinfo{volume}{106}}, \bibinfo{pages}{012409},
  \doiprefix\url{10.1103/PhysRevA.106.012409} (\bibinfo{year}{2022}).
\newblock \eprint{2206.06736}.

\bibitem{Quek2018A1812.06693}
\bibinfo{author}{{Quek}, Y.}, \bibinfo{author}{{Fort}, S.} \&
  \bibinfo{author}{{Khoon Ng}, H.}
\newblock \bibinfo{journal}{\bibinfo{title}{{Adaptive Quantum State Tomography
  with Neural Networks}}}.
\newblock {\emph{\JournalTitle{npj Quantum Information}}}
  \textbf{\bibinfo{volume}{7}}, \bibinfo{pages}{105},
  \doiprefix\url{10.1038/s41534-021-00436-9} (\bibinfo{year}{2021}).
\newblock \eprint{1812.06693}.

\bibitem{Iouchtchenko2023N2206.15449}
\bibinfo{author}{{Iouchtchenko}, D.}, \bibinfo{author}{{Gonthier}, J.~F.},
  \bibinfo{author}{{Perdomo-Ortiz}, A.} \& \bibinfo{author}{{Melko}, R.~G.}
\newblock \bibinfo{journal}{\bibinfo{title}{{Neural network enhanced
  measurement efficiency for molecular groundstates}}}.
\newblock {\emph{\JournalTitle{Machine Learning: Science and Technology}}}
  \textbf{\bibinfo{volume}{4}}, \bibinfo{pages}{015016},
  \doiprefix\url{10.1088/2632-2153/acb4df} (\bibinfo{year}{2023}).
\newblock \eprint{2206.15449}.

\bibitem{Carrasquilla2018R1810.10584}
\bibinfo{author}{{Carrasquilla}, J.}, \bibinfo{author}{{Torlai}, G.},
  \bibinfo{author}{{Melko}, R.~G.} \& \bibinfo{author}{{Aolita}, L.}
\newblock \bibinfo{journal}{\bibinfo{title}{{Reconstructing quantum states with
  generative models}}}.
\newblock {\emph{\JournalTitle{Nature Machine Intelligence}}}
  \textbf{\bibinfo{volume}{1}}, \bibinfo{pages}{155--161},
  \doiprefix\url{10.1038/s42256-019-0028-1} (\bibinfo{year}{2019}).
\newblock \eprint{1810.10584}.

\bibitem{Carrasquilla2021P1912.11052}
\bibinfo{author}{{Carrasquilla}, J.} \emph{et~al.}
\newblock \bibinfo{journal}{\bibinfo{title}{{Probabilistic simulation of
  quantum circuits using a deep-learning architecture}}}.
\newblock {\emph{\JournalTitle{Physical Review A}}}
  \textbf{\bibinfo{volume}{104}}, \bibinfo{pages}{032610},
  \doiprefix\url{10.1103/PhysRevA.104.032610} (\bibinfo{year}{2021}).
\newblock \eprint{1912.11052}.

\bibitem{Cha2022A2006.12469}
\bibinfo{author}{{Cha}, P.} \emph{et~al.}
\newblock \bibinfo{journal}{\bibinfo{title}{{Attention-based quantum
  tomography}}}.
\newblock {\emph{\JournalTitle{Machine Learning: Science and Technology}}}
  \textbf{\bibinfo{volume}{3}}, \bibinfo{pages}{01LT01},
  \doiprefix\url{10.1088/2632-2153/ac362b} (\bibinfo{year}{2022}).
\newblock \eprint{2006.12469}.

\bibitem{Goldt2020T2006.14709}
\bibinfo{author}{{Goldt}, S.} \emph{et~al.}
\newblock \bibinfo{journal}{\bibinfo{title}{{The Gaussian equivalence of
  generative models for learning with shallow neural networks}}}.
\newblock {\emph{\JournalTitle{arXiv e-prints}}}
  \bibinfo{pages}{arXiv:2006.14709}, \doiprefix\url{10.48550/arXiv.2006.14709}
  (\bibinfo{year}{2020}).
\newblock \eprint{2006.14709}.

\bibitem{Ingrosso2022D2202.00565}
\bibinfo{author}{{Ingrosso}, A.} \& \bibinfo{author}{{Goldt}, S.}
\newblock \bibinfo{journal}{\bibinfo{title}{{Data-driven emergence of
  convolutional structure in neural networks}}}.
\newblock {\emph{\JournalTitle{Proceedings of the National Academy of
  Science}}} \textbf{\bibinfo{volume}{119}}, \bibinfo{pages}{e2201854119},
  \doiprefix\url{10.1073/pnas.2201854119} (\bibinfo{year}{2022}).
\newblock \eprint{2202.00565}.

\bibitem{Refinetti2022N2211.11567}
\bibinfo{author}{{Refinetti}, M.}, \bibinfo{author}{{Ingrosso}, A.} \&
  \bibinfo{author}{{Goldt}, S.}
\newblock \bibinfo{journal}{\bibinfo{title}{{Neural networks trained with SGD
  learn distributions of increasing complexity}}}.
\newblock {\emph{\JournalTitle{arXiv e-prints}}}
  \bibinfo{pages}{arXiv:2211.11567}, \doiprefix\url{10.48550/arXiv.2211.11567}
  (\bibinfo{year}{2022}).
\newblock \eprint{2211.11567}.

\bibitem{Iten2020D1807.10300}
\bibinfo{author}{{Iten}, R.}, \bibinfo{author}{{Metger}, T.},
  \bibinfo{author}{{Wilming}, H.}, \bibinfo{author}{{del Rio}, L.} \&
  \bibinfo{author}{{Renner}, R.}
\newblock \bibinfo{journal}{\bibinfo{title}{{Discovering Physical Concepts with
  Neural Networks}}}.
\newblock {\emph{\JournalTitle{Physical Review Letters}}}
  \textbf{\bibinfo{volume}{124}}, \bibinfo{pages}{010508},
  \doiprefix\url{10.1103/PhysRevLett.124.010508} (\bibinfo{year}{2020}).
\newblock \eprint{1807.10300}.

\bibitem{Poulsen-Nautrup2020O2001.00593}
\bibinfo{author}{{Poulsen Nautrup}, H.} \emph{et~al.}
\newblock \bibinfo{journal}{\bibinfo{title}{{Operationally meaningful
  representations of physical systems in neural networks}}}.
\newblock {\emph{\JournalTitle{arXiv e-prints}}}
  \bibinfo{pages}{arXiv:2001.00593}, \doiprefix\url{10.48550/arXiv.2001.00593}
  (\bibinfo{year}{2020}).
\newblock \eprint{2001.00593}.

\bibitem{Frohnert2023E2306.05694}
\bibinfo{author}{{Frohnert}, F.} \& \bibinfo{author}{{van Nieuwenburg}, E.}
\newblock \bibinfo{journal}{\bibinfo{title}{{Explainable Representation
  Learning of Small Quantum States}}}.
\newblock {\emph{\JournalTitle{arXiv e-prints}}}
  \bibinfo{pages}{arXiv:2306.05694}, \doiprefix\url{10.48550/arXiv.2306.05694}
  (\bibinfo{year}{2023}).
\newblock \eprint{2306.05694}.

\bibitem{Preskill2018Q1801.00862}
\bibinfo{author}{{Preskill}, J.}
\newblock \bibinfo{journal}{\bibinfo{title}{{Quantum Computing in the NISQ era
  and beyond}}}.
\newblock {\emph{\JournalTitle{Quantum}}} \textbf{\bibinfo{volume}{2}},
  \bibinfo{pages}{79}, \doiprefix\url{10.22331/q-2018-08-06-79}
  (\bibinfo{year}{2018}).
\newblock \eprint{1801.00862}.

\bibitem{Hu2022H2102.10132}
\bibinfo{author}{{Hu}, H.-Y.} \& \bibinfo{author}{{You}, Y.-Z.}
\newblock \bibinfo{journal}{\bibinfo{title}{{Hamiltonian-driven shadow
  tomography of quantum states}}}.
\newblock {\emph{\JournalTitle{Physical Review Research}}}
  \textbf{\bibinfo{volume}{4}}, \bibinfo{pages}{013054},
  \doiprefix\url{10.1103/PhysRevResearch.4.013054} (\bibinfo{year}{2022}).
\newblock \eprint{2102.10132}.

\bibitem{Hu2023C2107.04817}
\bibinfo{author}{{Hu}, H.-Y.}, \bibinfo{author}{{Choi}, S.} \&
  \bibinfo{author}{{You}, Y.-Z.}
\newblock \bibinfo{journal}{\bibinfo{title}{{Classical Shadow Tomography with
  Locally Scrambled Quantum Dynamics}}}.
\newblock {\emph{\JournalTitle{Physical Review Research}}}
  \textbf{\bibinfo{volume}{5}}, \bibinfo{pages}{023027},
  \doiprefix\url{10.1103/PhysRevResearch.5.023027} (\bibinfo{year}{2023}).
\newblock \eprint{2107.04817}.

\bibitem{Akhtar2023S2209.02093}
\bibinfo{author}{{Akhtar}, A.~A.}, \bibinfo{author}{{Hu}, H.-Y.} \&
  \bibinfo{author}{{You}, Y.-Z.}
\newblock \bibinfo{journal}{\bibinfo{title}{{Scalable and Flexible Classical
  Shadow Tomography with Tensor Networks}}}.
\newblock {\emph{\JournalTitle{Quantum}}} \textbf{\bibinfo{volume}{7}},
  \bibinfo{pages}{1026}, \doiprefix\url{10.22331/q-2023-06-01-1026}
  (\bibinfo{year}{2023}).
\newblock \eprint{2209.02093}.

\bibitem{Bertoni2022S2209.12924}
\bibinfo{author}{{Bertoni}, C.} \emph{et~al.}
\newblock \bibinfo{journal}{\bibinfo{title}{{Shallow shadows: Expectation
  estimation using low-depth random Clifford circuits}}}.
\newblock {\emph{\JournalTitle{arXiv e-prints}}}
  \bibinfo{pages}{arXiv:2209.12924} (\bibinfo{year}{2022}).
\newblock \eprint{2209.12924}.

\bibitem{Ippoliti2023O2212.11963}
\bibinfo{author}{{Ippoliti}, M.}, \bibinfo{author}{{Li}, Y.},
  \bibinfo{author}{{Rakovszky}, T.} \& \bibinfo{author}{{Khemani}, V.}
\newblock \bibinfo{journal}{\bibinfo{title}{{Operator Relaxation and the
  Optimal Depth of Classical Shadows}}}.
\newblock {\emph{\JournalTitle{Physical Review Letters}}}
  \textbf{\bibinfo{volume}{130}}, \bibinfo{pages}{230403},
  \doiprefix\url{10.1103/PhysRevLett.130.230403} (\bibinfo{year}{2023}).
\newblock \eprint{2212.11963}.

\end{thebibliography}



\section*{Acknowledgements}

We acknowledge the helpful discussions with Xiao-Liang Qi, Hong-Ye Hu, Roger Melko, and John McGreevy. The research project is supported by Y.Z.Y.'s personal fund. We thank Lambda Labs for the GPU cloud service. We are grateful to ChatGPT for providing linguistic advice in the writing of this article.

\section*{Author contributions statement}

Y.Z.Y. conceived and led the research project. Z.Z. and Y.Z.Y. developed the algorithm, trained the models, collected the data, and analyzed the results. All authors drafted and reviewed the manuscript. 

\section*{Supplementary Information}

\subsection*{Samples of Classical Shadows}

For demonstration purposes, we list 30 samples of classical shadows $(\vect{x},\vect{y})$ collected from randomized Pauli measurement on the $N=5$ GHZ state. They resemble the classical noise in the environment after the decoherence of Schrödinger's cat and encode the classical information about the original quantum state of the cat.

\eqs{
& \begin{array}{ll}\vect{x}:& \ttt{XXYXZ}\\ \vect{y}:& \ttt{---+-} \end{array} \quad \begin{array}{ll}\vect{x}:& \ttt{XYYYY}\\ \vect{y}:& \ttt{-++-+} \end{array} \quad \begin{array}{ll}\vect{x}:& \ttt{XYXYZ}\\ \vect{y}:& \ttt{+----} \end{array} \quad \begin{array}{ll}\vect{x}:& \ttt{XZXXY}\\ \vect{y}:& \ttt{+-+--} \end{array} \quad \begin{array}{ll}\vect{x}:& \ttt{YXXXX}\\ \vect{y}:& \ttt{--++-} \end{array} \quad \begin{array}{ll}\vect{x}:& \ttt{YZZZX}\\ \vect{y}:& \ttt{----+} \end{array} \\ & \begin{array}{ll}\vect{x}:& \ttt{YZYZY}\\ \vect{y}:& \ttt{-+++-} \end{array} \quad \begin{array}{ll}\vect{x}:& \ttt{YXYXZ}\\ \vect{y}:& \ttt{+-+++} \end{array} \quad \begin{array}{ll}\vect{x}:& \ttt{YYZZY}\\ \vect{y}:& \ttt{+-++-} \end{array} \quad \begin{array}{ll}\vect{x}:& \ttt{YZYYX}\\ \vect{y}:& \ttt{+-++-} \end{array} \quad \begin{array}{ll}\vect{x}:& \ttt{YZZXZ}\\ \vect{y}:& \ttt{+--+-} \end{array} \quad \begin{array}{ll}\vect{x}:& \ttt{YZYYX}\\ \vect{y}:& \ttt{++-++} \end{array} \\ & \begin{array}{ll}\vect{x}:& \ttt{YZYXX}\\ \vect{y}:& \ttt{+++++} \end{array} \quad \begin{array}{ll}\vect{x}:& \ttt{ZXXZX}\\ \vect{y}:& \ttt{--+--} \end{array} \quad \begin{array}{ll}\vect{x}:& \ttt{ZXXZZ}\\ \vect{y}:& \ttt{--+--} \end{array} \quad \begin{array}{ll}\vect{x}:& \ttt{ZXXYX}\\ \vect{y}:& \ttt{-++-+} \end{array} \quad \begin{array}{ll}\vect{x}:& \ttt{ZXZZZ}\\ \vect{y}:& \ttt{-+---} \end{array} \quad \begin{array}{ll}\vect{x}:& \ttt{ZYXXZ}\\ \vect{y}:& \ttt{--+--} \end{array} \\ & \begin{array}{ll}\vect{x}:& \ttt{ZYXXX}\\ \vect{y}:& \ttt{--+++} \end{array} \quad \begin{array}{ll}\vect{x}:& \ttt{ZXXZZ}\\ \vect{y}:& \ttt{++-++} \end{array} \quad \begin{array}{ll}\vect{x}:& \ttt{ZYXXX}\\ \vect{y}:& \ttt{+--++} \end{array} \quad \begin{array}{ll}\vect{x}:& \ttt{ZYXYZ}\\ \vect{y}:& \ttt{+-+++} \end{array} \quad \begin{array}{ll}\vect{x}:& \ttt{ZYXYY}\\ \vect{y}:& \ttt{++--+} \end{array} \quad \begin{array}{ll}\vect{x}:& \ttt{ZYYXY}\\ \vect{y}:& \ttt{++--+} \end{array} \\ & \begin{array}{ll}\vect{x}:& \ttt{ZYYXX}\\ \vect{y}:& \ttt{++-+-} \end{array} \quad \begin{array}{ll}\vect{x}:& \ttt{ZZXXY}\\ \vect{y}:& \ttt{+++--} \end{array} \quad \begin{array}{ll}\vect{x}:& \ttt{ZZXXX}\\ \vect{y}:& \ttt{++++-} \end{array} \quad \begin{array}{ll}\vect{x}:& \ttt{ZZYZX}\\ \vect{y}:& \ttt{++-+-} \end{array} \quad \begin{array}{ll}\vect{x}:& \ttt{ZZYXY}\\ \vect{y}:& \ttt{+++--} \end{array} \quad \begin{array}{ll}\vect{x}:& \ttt{ZZZYZ}\\ \vect{y}:& \ttt{+++-+} \end{array} \\  & \cdots
}

\subsection*{Model Architecture}

\begin{figure}[htbp]
\begin{center}
\includegraphics[scale=0.7]{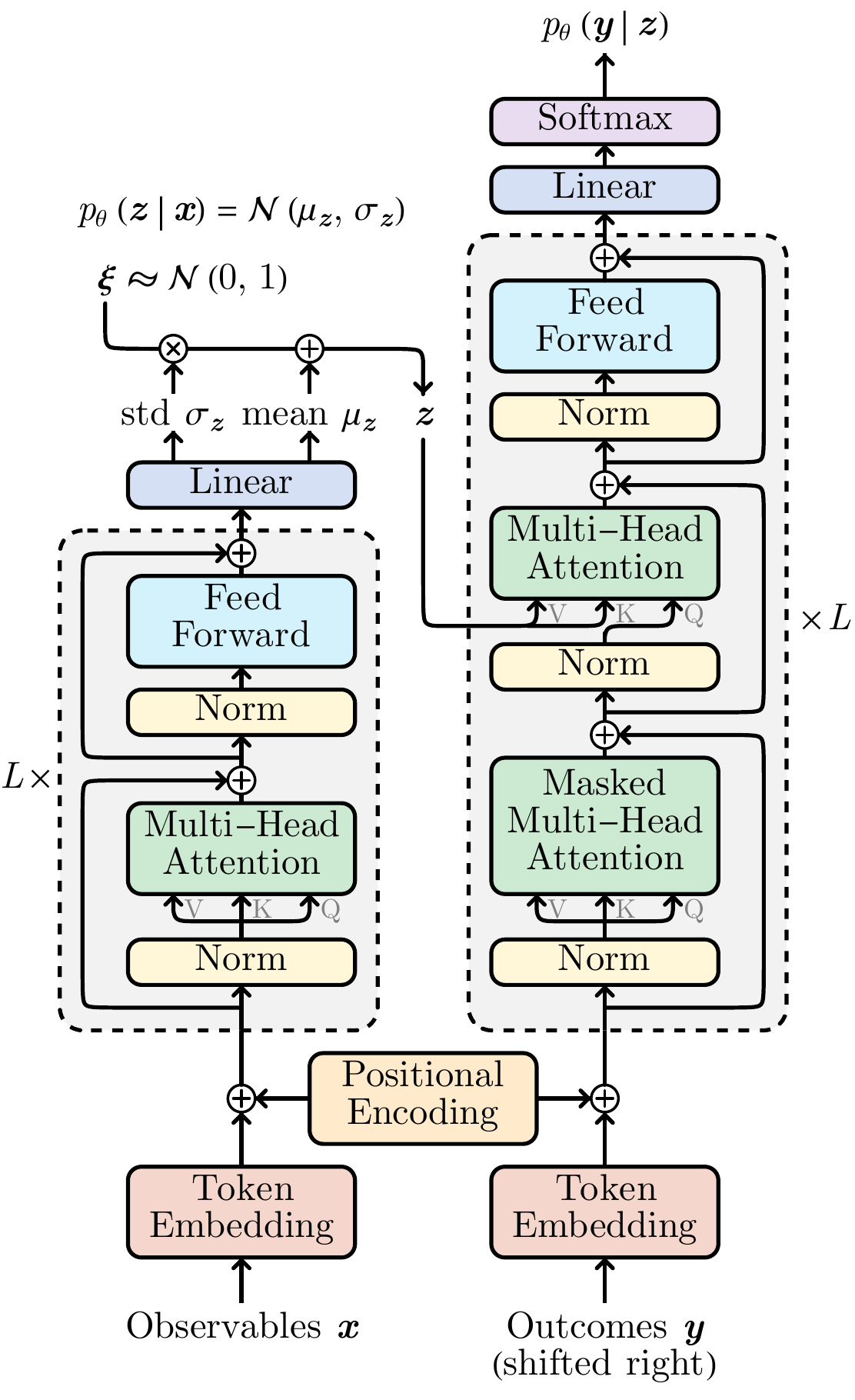}
\caption{The transformer-based $\beta$-VAE architecture.}
\label{fig: transformer}
\end{center}
\end{figure}

The model architecture is illustrated in \figref{fig: transformer}. 

On the encoder side, the length-$N$ observable sequence $\vect{x}$ is first embedded as $N$ vectors in a $d$-dimensional embedding space. Each vector is then dressed by positional encodings. The encoder $L$ transformer encoding layers to map these input vectors (shape $N\times d$) to the mean $\mu_\vect{z}$ (shape $N\times d$) and the (diagonal) standard deviation $\sigma_\vect{z}$ (shape $N\times d$) of latent variables. Then the probability $p_\theta(\vect{z}|\vect{x})$ is modeled as a Gaussian distribution
\eq{p_\theta(\vect{z}|\vect{x})=\frac{1}{(2\pi)^{Nd/2}\det\sigma_\vect{z}}\exp\Big(-\frac{(\vect{z}-\mu_\vect{z})^2}{2\sigma_\vect{z}^2}\Big).}
One can adopt the reparametrization trick to sample from the Gaussian distribution by first sampling $\vect{\xi}$ from the standard normal distribution $\scN(0,1)$, and then construct $\vect{z}=\sigma_\vect{z}\vect{\xi}+\mu_\vect{z}$. This allows the gradient to back-propagate through the sampling step.

On the decoder side, the length-$N$ measurement outcome sequence $\vect{y}$ is first shifted right by one token (with a null token prepended). The tokens are then embedded as $N$ vectors of dimension $d$ and dressed by the positional encodings as well. These vectors (shape $N\times d$) are combined with the latent variables $\vect{z}$ (shape $N\times d$) by the decoder through $L$ transformer decoding layers. The end result (shape $N\times d$) is projected by a linear layer to 2-dimensional vectors (shape $N\times 2$). After softmax, the vector components represent the conditional probabilities $p_\theta(y_i=\pm|\vect{y}_{<i},\vect{z})$ for $i=1,2,\cdots,N$. In this way, the probability $p_\theta(\vect{y}|\vect{z})$ is modeled by
\eq{p_\theta(\vect{y}|\vect{z})=\prod_{i=1}^{N}p_\theta(y_i|\vect{y}_{<i},\vect{z}).}
The sampling can be performed autoregressively.

The hyperparameters of our model are set to $(d,L)=(64,1)$ for $N=1,2,\cdots,5$ and $(d,L)=(128,2)$ for $N=6$. The number of attention heads is always 16.

\subsection*{Reconstructed Density Matrices}

\begin{figure}[htbp]
\begin{center}
\includegraphics[scale=0.7]{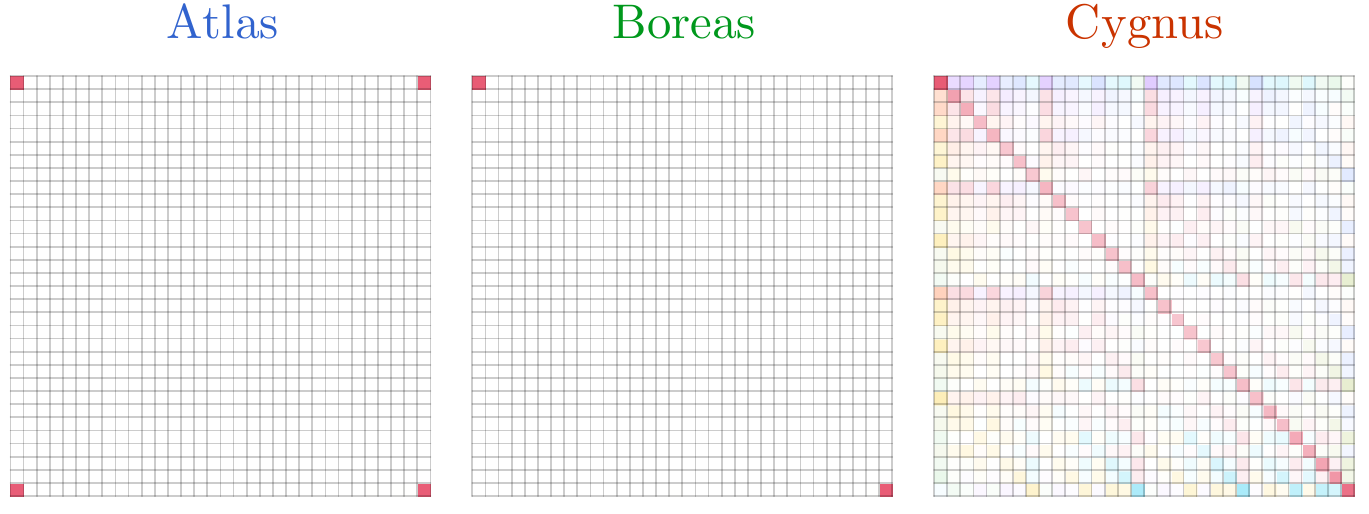}
\caption{The reconstructed quantum states ($32\times 32$ density matrices) $\rho_\text{mdl}$ based on the classical shadows generated by the three representative models respectively. Each pixel represents a matrix element. }
\label{fig: rhos}
\end{center}
\end{figure}

We can sample classical shadows from the trained generative language models and reconstruct the density matrix $\rho_\text{mdl}$ using the reconstruction formula in \eqnref{eq: rec mdl}. \figref{fig: rhos} presents the visualizations of these density matrices in the computational basis (the $Z$-basis), as reconstructed by Atlas, Boreas and Cygnus respectively. Darker pixel represents larger matrix elements, and the color encodes the complex phase ($+1$: red, $+\ii$: yellow, $-1$: green, $-\ii$: blue).

Atlas correctly reconstructs the full quantum density matrix of the GHZ state $\ket{0}^{\otimes N}+\ket{1}^{\otimes N}$. Boreas fails to capture the off-diagonal matrix elements that represent quantum coherence, as a result, the density matrix is decoherent. Nevertheless, Boreas correctly captures the two classical states ($\ket{0}^{\otimes N}$ and $\ket{1}^{\otimes N}$) represented by the two diagonal matrix elements. Cygnus's reconstructed density matrix is close to an identity matrix (with noisy off-diagonal patterns), indicating that it has not event realized the two classical realities of the cat state.

\section*{Additional information}

\subsection*{Data Availability}
The source code and the datasets generated and analyzed during the current study are available in the corresponding GitHub repository \url{https://github.com/EverettYou/EmergentClassicality}.

\subsection*{Competing Interests}
The authors declare that they have no competing interests.

\end{document}